\documentclass[aps,twocolumn,showpacs,showkeys,preprintnumbers,amsmath,amssymb,floatfix]{revtex4}

\usepackage{graphicx}
\usepackage{epsfig}
\usepackage{rotate}

\begin{document}
\title{Fission Decay of N=Z nuclei at high angular momentum:  $^{60}$Zn
                 }
\author{W.~von~Oertzen$^{1,2}$} 
\email{oertzen@hmi.de}
\author{V. Zherebchevsky$^{1,3}$} 
\author{B.~Gebauer$^1$} 
\author{Ch.~Schulz$^1$} 
\author{S.~Thummerer$^1$} 
\author{D. Kamanin$^4$} 
\author{G.~Royer$^5$} 
\author{Th.~Wilpert$^1$}

\noindent
\address{$^1$ Hahn-Meitner-Institut-GmbH, Glienicker Strasse 100, 
D-14109 Berlin, Germany}
\address{$^2$ Fachbereich Physik, Free University of Berlin, Germany}
\address{$^3$ St.Petersburg University, St.Petersburg, Russia}
\address{$^4$ FLNR, JINR, Dubna, Russia} 
\address{$^5$ Laboratoire Subatech, Universit\'e-IN2P3/CNRS-Ecole des Mines,
                44307 Nantes, France}

\date{\today}

\begin{abstract}
Using a unique two-arm detector system for heavy ions 
(the BRS, binary reaction spectrometer) 
coincident fission events have been measured 
from the decay of $^{60}$Zn compound nuclei formed at 88 MeV 
excitation energy in the reactions with 
$^{36}$Ar beams on a $^{24}$Mg target at
E$_{lab}$($^{36}$Ar) =  195 MeV.
The detectors consisted of two large area position sensitive (x,y) gas
telescopes with Bragg-ionization chambers.
From the  binary 
coincidences in the two detectors inclusive and exclusive cross sections for 
fission channels with differing losses of charge
were obtained. Narrow out-of-plane correlations corresponding to 
coplanar decay are observed for
two fragments emitted in binary events, 
and in the data for ternary decay with missing charges from 4 up to 8.
 After subtraction 
of broad components
these narrow correlations are interpreted
 as a ternary fission process at high angular momentum through an elongated
shape. The lighter mass in the neck region consists dominantly of two or three 
$\alpha$-particles. Differential cross sections for the different mass splits 
for binary and ternary fission are presented.
The relative yields of the binary and ternary events  are 
explained using the statistical model based on the extended 
Hauser-Feshbach formalism for compound nucleus decay.
The ternary fission process can be described by the decay of 
hyper-deformed states with angular momentum around 45-52 $\hbar$. 
\end{abstract}

\pacs{ 25.70.Jj, 25.70.Pq, 24.60.Dr, 23.70.+j} 
\keywords{$^{60}$Zn compound nucleus formation from $^{36}$Ar+$^{24}$Mg,
          binary and ternary fission, coincidence measurement technique. }

\maketitle

\section{Introduction}
\label{intro}

For nuclei with masses ranging from A = 40 up to 100  at high angular momentum 
 "exotic"  shapes called super- and hyper-deformed are predicted 
within various theoretical 
approaches~\cite{swiatecki,moeller76,zhang94,lea75,aberg90,aberg94,Rag81}. For these nuclei, 
fission decay 
is expected to proceed through  shapes with  large deformations in particular 
for the  large angular momenta. 
Some authors have  indicated that the structure of the fissioning nucleus at the saddle point 
is strongly determined by clustering~\cite{zhang94,aberg90}, 
which can influence the fission decay modes with cluster emission.
Studies of the binary 
decays, as well as cluster emission  of such compound nuclei (CN)
 in the mass region A = 40 - 60, have
been reported in the last 
two decades~\cite{sanders94,sanders87,sanders89,sanders99,kokalova05}.

 The most important factors governing the fission decay
can be described by the two fissility parameters $\eta_i$, which  
determine the neck formation~\cite{swiatecki}. The parameter $\eta_x$ 
is used as a 
measure of the relative strength of two competing forces due to 
 the Coulomb energy (repulsive) and the surface energy
(attractive), and further the parameter $\eta_y$ is introduced to describe the 
relative strength of the centrifugal energy  (repulsive) determined by
the angular momentum $L$ and  the surface energy (attractive).
For nuclei of mass $A$, charge $Z$, and radius $R$, these parameters are given by 
\begin{equation}
  \label{eq:eta}
  \eta_{x} = \frac{E_{Coul}}{2E_{Surf}} = C_x\frac{Z^2}{A},  \\ 
  \eta_{y} = \frac{E_{Rot}}{E_{Surf}} = C_y\frac{L^2}{AR^4}, 
\end{equation}
with two constants $ C_x$ and $ C_y$, respectively.
For large values of the fissility parameter, $\eta_x$ = 0.7 - 0.8, in heavy nuclei, 
the saddle point configurations are predicted~\cite{swiatecki} to be
very compact, no real neck is formed. The quadrupole deformation parameter
$\beta_2$ at the saddle point may be in the range of only 0.2 - 0.3. In contrast,
for small values of the fissility parameter, $\eta_x$=0.3 - 0.4, for 
light nuclei as in our case,
the configurations of the saddle point are predicted to become very elongated
~\cite{swiatecki,moeller76} with decreasing fission parameter.
For nuclei in the mass region of A = 56 - 100   a ternary fission process  
is expected to occur.
Further the neck becomes longer and ``thicker'' at higher angular momenta, 
 i.e. with increasing values of the second fissility parameter
$\eta_{y}$. For the excited states in light N=Z nuclei at high angular
momentum super-deformed and very elongated hyper-deformed shapes 
are predicted, stabilised 
by special quantal shell effects~\cite{zhang94,lea75,aberg90,aberg94,Rag81}. 
In these nuclei (A = 40-100) energetically favoured states with super- and
hyper-deformed shapes with quadrupole deformation parameters $\beta_2$ = 0.6 - 1.0 
(with 2:1 and 3:1 as ratios between the major to minor axis for  
ellipsoidal deformation, respectively) and even larger values are predicted with
the Nilsson-Strutinsky method~\cite{aberg90,aberg94,Rag81}. 
The well established  shell corrections will 
stabilise the rotating CN in its hyper-deformed shape at 
high angular momentum~\cite{aberg94}. Alternatively hyper-deformed configurations 
are also obtained in an $\alpha$-cluster model~\cite{zhang94}, a fact 
which  gives strong indications
for the relation between  large deformations and clustering.
From these results it can be anticipated that in the mass region of the
present work (A = 50-60) the
super- and hyper-deformed states at high angular momentum will have fission  as 
one of the dominant decay modes~\cite{sanders99,Andreev06}).

\begin{figure}[htbp]
  \begin{center}
      \hspace{-3mm} 
     \includegraphics[width=0.50\textwidth]{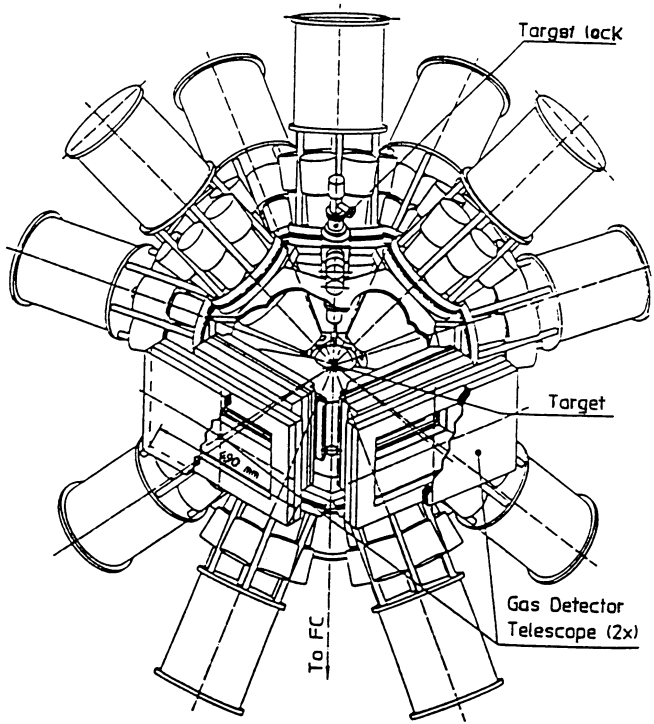}
    \caption{The BRS-OSIRIS set-up used
     for the study of coincident fission channels 
     in the reactions  $^{36}$Ar + $^{24}$Mg$\rightarrow$  
     $Z_3+Z_4$+$\Delta Z$, at E$_{lab}$($^{36}$Ar) = 195 MeV.
   The OSIRIS $\gamma$-detector array consisted of 12 Compton-suppressed Ge-detectors.}
    \label{fig:BRS}
  \end{center}
\end{figure}

\begin{figure*}[htbp]
  \begin{center}
    \includegraphics[width=0.85\textwidth]{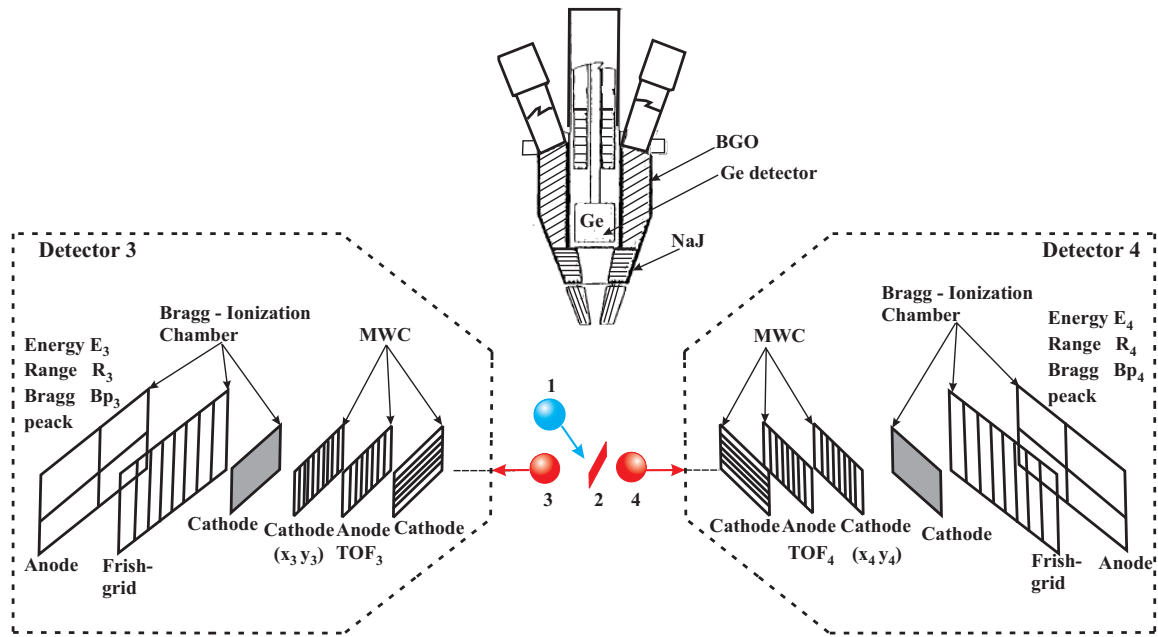}
    \caption{(Color online) Schematic representation of the telescopes of the
  Binary Reaction Spectrometer (BRS), they  are shown
  together with one  cluster-$\gamma$-detector of the OSIRIS-array.
  The position sensitive (x,y) low-pressure multi-wire chamber (MWC)
  and  part and the Bragg-Ionization Chamber (BIC) are indicated.}
    \label{fig:dect}
 \end{center}
\end{figure*}

\begin{figure}[htbp]
  \begin{center}
    \includegraphics[width=0.48\textwidth]{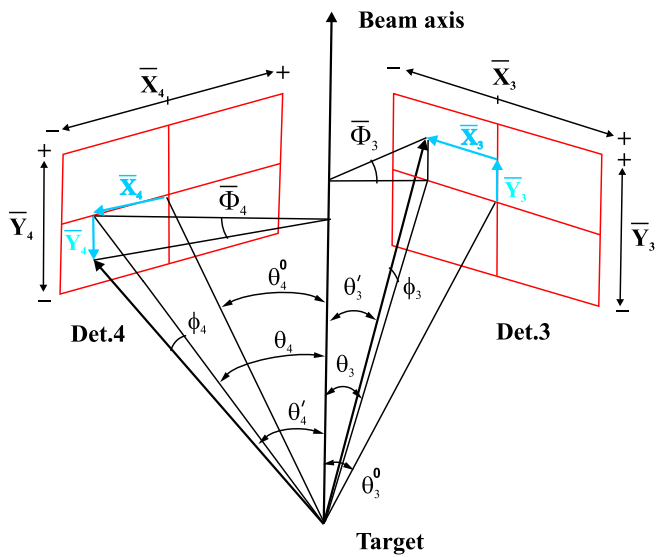}
    \caption{ (Color online)
   Definitions of the in-plane ($\theta_3$, $\theta_4$) and out-of-plane
  ($\phi_3$, $\phi_4$) scattering angles obtained from the
  $\overline{X}$ and $\overline{Y}$ read-outs 
   of the position sensitive detectors.}
    \label{fig:dect1}
\end{center}
\end{figure}
The  ternary fission process is strongly enhanced for the configurations 
with the largest deformations due to
the large moments of inertia and the corresponding 
lowering of the fission barrier, which in addition can be decreased 
by the mentioned shell corrections. In another approach, using a generalised 
 liquid drop model, taking into account the proximity energy and 
quasi-molecular shapes (as in the cluster models)
 Royer et al.~\cite{roy95,royer95} have also predicted  several of 
such ternary fission modes in 
$^{56}$Ni and $^{48}$Cr compound nuclei. In these systems even macroscopically
 super- and hyper-deformed minima appear at the foot of the potential fission 
barrier  at sufficiently high angular momenta. However, until now experimental 
evidence for such ternary breakup of the light  nuclei has not 
been reported (e.g. ref.~\cite{murphy96}). Previous reports on the 
results of the present work are 
contained in~\cite{thummerer98,zhereb06,zhereb07,Efimov07}, and the most recent work
 on $^{56}$Ni in ref.~\cite{voe08}. 
Work on fission in a similar mass range has 
been  described in~\cite{sanders87,sanders89,sanders99}.  Ternary 
 fission  in the statistical approach has been  discussed  
in a recent study also for $^{56}$Ni~\cite{Andreev06}.

\begin{figure*}[htbp]
\begin{center}
 \includegraphics[width=0.99\textwidth]{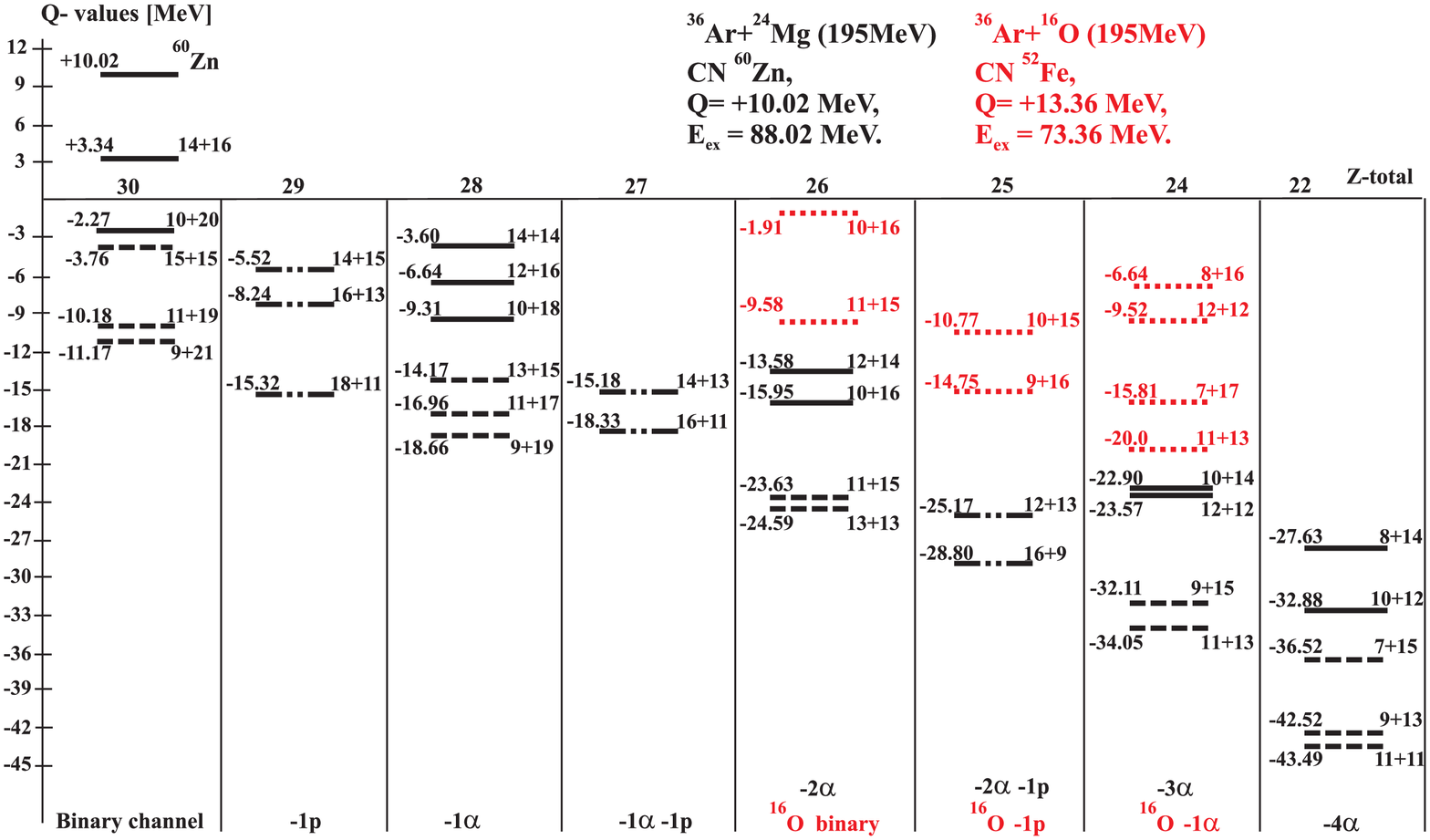}
\caption{ (Color online)
 Q-values for different binary and non-binary decay channels for
 the compound nucleus $^{60}$Zn and $^{52}$Fe, defined by the 
  charges  $Z3$ and $Z4$ ($Z3$+$Z4$) as
  indicated. The values are given for the even missing charges
  (missing number of $\alpha$-particles) and for odd missing charges
  (missing number of protons and p+x$\alpha$-particles)
   as indicated in the bottom line.}   
    \label{fig:Qvalue_60Zn}
\end{center}
\end{figure*}

\section{The detector system BRS}
\label{detec}
The present experiment has been  performed with the binary reaction
 spectrometer (BRS) in combination with the $\gamma$-detector array, OSIRIS, 
at the HMI (Berlin) with a $^{36}$Ar beam of 195 MeV~\cite{ref:gebauer98,beck04}.
A similar experiment has been performed  with a $^{32}$S beam 
in the BRS-EUROBALL-IV campaign~\cite{Efimov07}.
The experimental set-up is  shown in Fig.~\ref{fig:BRS}.
In these experiments coincidences of $\gamma$-rays (with the OSIRIS-array) with charged 
fragments have been used for $\gamma$-spectroscopy.  Two heavy fragments are 
registered in kinematical coincidence and identified by their charge in two
large area position sensitive Bragg-Ionization Chambers (BIC)~\cite{ref:gebauer93}.
Two detector telescopes labelled 3 and 4 (Det3 and Det4) are
placed symmetrically to the beam axis and consist of two-dimensional
position-sensitive low-pressure multi-wire chambers (MWC) and Bragg-curve
ionization chambers. 
All detection planes are electrically 
fourfold segmented, in order to improve the resolution and counting rate capabilities.
 A scheme is shown in  Fig.\ref{fig:dect},
there the arrangement of the position sensitive  part and the Bragg-ionization
chambers are shown.  The parameters giving the
identification of the fragments by their charge are the
Bragg-peak height $BP$, the range $R$, and the rest energy $E$. Their momentum vectors are obtained 
through the position signals ($x$ and $y$)  of the two telescopes, from which the angles  
in the reaction plane, $\theta$, and  out-of-plane, $\phi$, are determined, 
 see Fig.\ref{fig:dect1}.

With the BRS it is therefore possible to measure two
heavy fragments in  coincidence with respect to their
in-plane ($\theta_{3},\theta_{4}$), and 
out-of-plane scattering angles ($\phi_{3}$, $\phi_{4}$),
time of flight (TOF) and energy (E).
In comparison to previous works these coincidences 
represent an exclusive measurement 
of the binary fission yield. Some more details of the detectors 
can be seen in Fig.~\ref{fig:dect}, see also~\cite{thummerer98}. 
The angular ranges of the two detectors cover a
very large solid angle of twice 156 msr ($\Delta\theta = 12^{\circ} - 46^{\circ}$,
$\Delta\phi=2\times 17.4^{\circ}$).
Examples of  $BP$ versus $E$ and $R$ spectra are shown
in Figs.~\ref{fig:bragg1} and ~\ref{fig:bragg2}.
The inclusive yields are obtained by setting just Z-gates in either one of the 
detectors.
For the exclusive detection of ``binary''
exit channels, two heavy fragments in coincidence are considered, with a definite choice
of  the charge  registered in Det3 and Det4. A survey of the reaction channels 
and their Q-values is given in Fig.~\ref{fig:Qvalue_60Zn}.
For the discussion of the reaction channels with two
heavy ejectiles the angular correlations with the  
in-plane, $\theta$, and out-of-plane, $\phi$, angles are the most relevant. 

\section{The experiment}
\label{experim}
\subsection{General experimental conditions}
\label{general}
The reaction studied in the present work can schematically be written as
 ($M_1,Z_1$)+($M_2,Z_2$) $\rightarrow$ $^{60}\textrm{Zn}^*$ $\rightarrow$ 
($M_3,Z_3$)+($\Delta$Z)+($M_4,Z_4$).
Two heavy fragments with masses $(M_3,M_4)$ and charges  $(Z_3,Z_4)$ are 
registered in kinematical coincidence and identified by their charges.
The masses could not be fully determined in
these experiments, however, a mass resolution of 3-4 mass units
was obtained by the timing signals 
of the MWC against the beam pulses (1-3 ns), 
see~\cite{thummerer98,zhereb07}.
We have investigated the fission decay of the
$^{60}$Zn CN  with the 
excitation energy of 88 MeV  and a maximum value of  the angular momentum
in the range of 44-52 $\hbar$.
A similar system, $^{32}$S+$^{24}$Mg, has been studied with the BRS-EUROBALL set-up 
and a detailed report is in preparation~\cite{Efimov07}.
The latter system has also been studied extensively by Sanders
et al.~\cite{sanders87,sanders89,sanders99}, with the emphasis on the
inclusive fusion cross section, the  binary
fission process  identified using kinematical coincidences.
 From this pioneering work some basic information for the CN
formation in the present reactions can be determined. For instance, the
maximum angular momentum reached for $^{56}$Ni, is close to 45$\hbar$, 
consistent with the predicted liquid drop limit~\cite{beck96}. At these
highest angular momenta the binary fission decay contributes up to 10 percent of
the total fusion cross section. It will be shown that the ternary fission decay 
in these nuclei competes with the binary fission due the large
 moments of inertia of the elongated shapes in the saddle point 
and the possible formation of  hyper-deformed configurations. 

\begin{figure}[htbp]
  \begin{center}
    \includegraphics[width=0.48\textwidth]{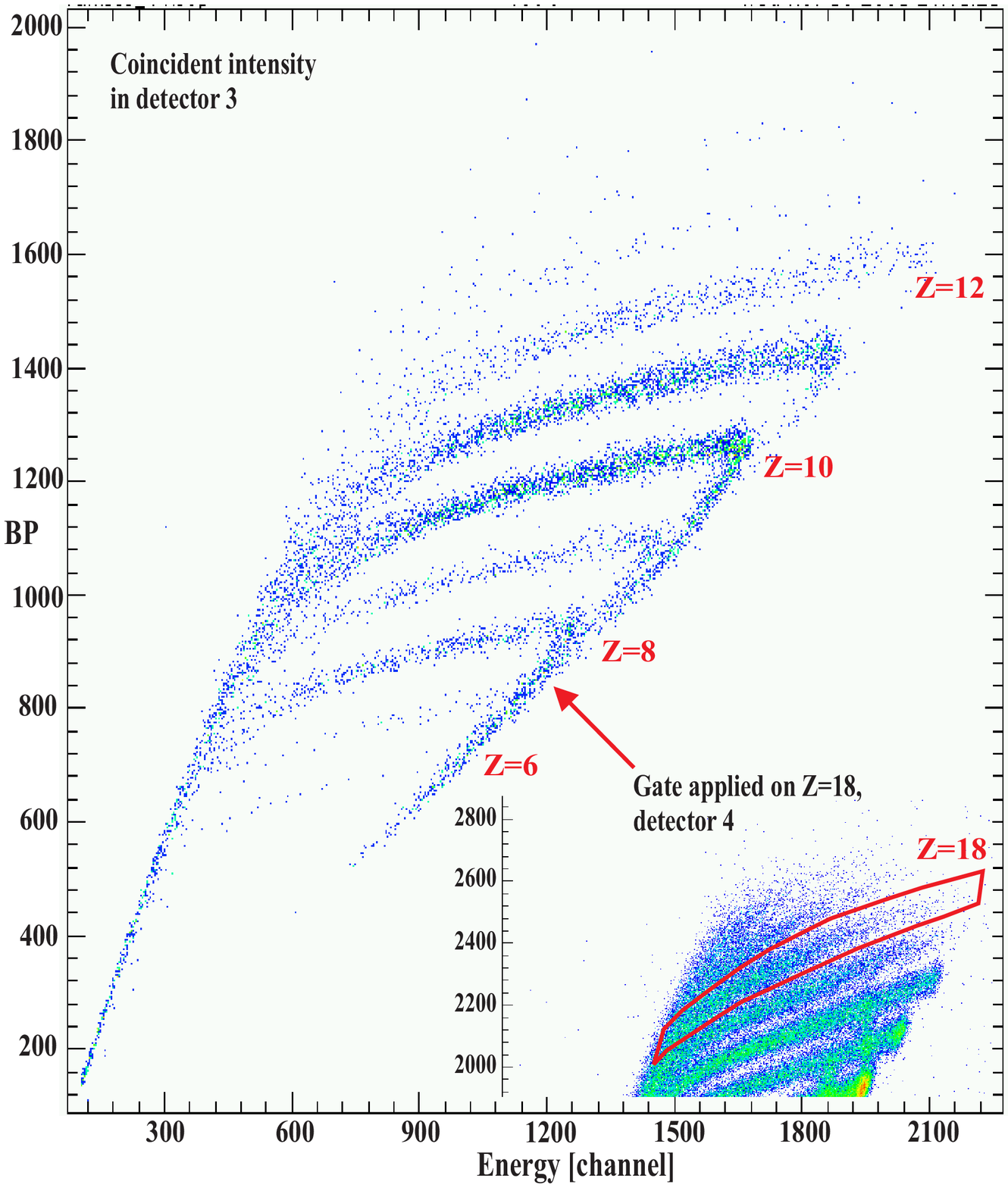}
    \caption{ (Color online)
             Bragg-peak (BP) - Energy distributions of the events in the
             detector 3, obtained in coincidence with a 
             gate set for a Z-value=18 in  detector 4 (insert), 
             for the reaction $^{36}$Ar+$^{24}$Mg$\rightarrow$  
                $Z_3+Z_4$+$\Delta Z$.}
    \label{fig:bragg1}
 \end{center}
\end{figure}

\begin{figure*}[htbp]
  \begin{center}
    \includegraphics[width=0.75\textwidth]{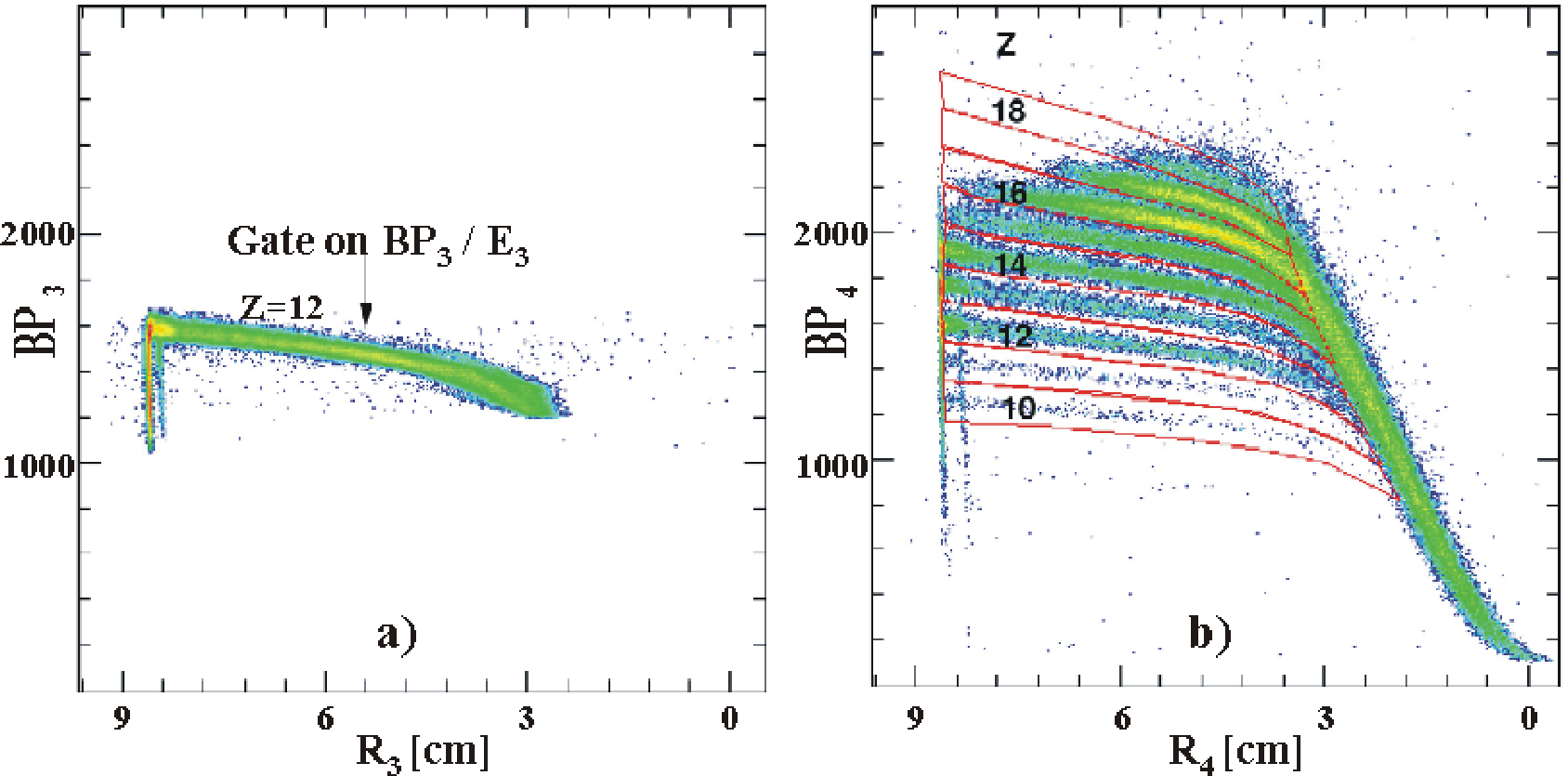}
    \caption{ (Color online)
            Bragg-peak (BP) versus  Range (R$_i$) distributions of the events in the
             Det.4, right side (b), obtained in coincidence  with a 
             gate set for Z-identification in the detector 3, left side 
            (a), for the reaction $^{36}$Ar+$^{24}$Mg$\rightarrow$  
                $Z_3+Z_4$+$\Delta Z$.}
    \label{fig:bragg2}
  \end{center}
\end{figure*}

\begin{figure}[htbp]
  \begin{center}
    \vspace{-7mm}
    \includegraphics[width=0.52\textwidth]{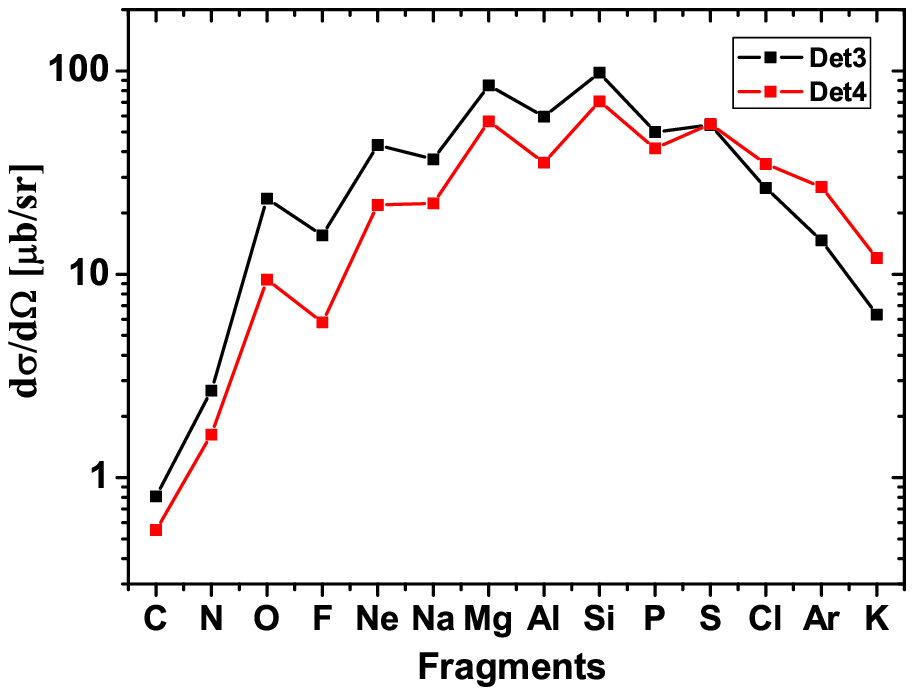}
    \vspace{-10mm}
    \caption{  (Color online)
 Inclusive coincidence yields, obtained with a 
 gate set for Z-identification in either of  the detectors (3) or (4)
 for the reaction $^{36}$Ar+$^{24}$Mg$\rightarrow$  
 $Z_3+ x$ or $ Z_4+x$, for an angular range of $\theta_{c.m.}=65^o - 120^o$.}
    \label{fig:inclAr}
   \end{center}
\end{figure}

We have observed  binary coincidences, channels defined by   $Z_3+Z_4$, 
with increasing loss of light particles, $\Delta Z$, 
($Z_{CN}-\Delta Z$=$Z_3+Z_4$), the
Q-values of these reactions are summarised in Fig.~\ref{fig:Qvalue_60Zn}.
The values become rather negative for larger $\Delta Z$.
We have selected binary and
non-binary channels by choosing a Z-gate in one detector (Det4, $Z_4$) and collecting
all coincident charge yields in the second detector (Det3, $Z_3$). The corresponding
yields in the Bragg-peak (BP) vs. Energy and the Bragg-peak vs. Range (R) distributions
are shown in Figs.~\ref{fig:bragg1} and \ref{fig:bragg2}, respectively.
This procedure gives the $inclusive$ binary yields, which  are shown in Fig.~\ref{fig:inclAr}. 

Choosing a second gate in Det3 we obtain the $exclusive$ reaction 
channels for the sum of the charges ($Z_3+Z_4$). This procedure defines channels with 
a fixed  missing charge,
 in the range $\Delta$Z = (0 - 8). 
These exclusive yields are the subject of the detailed discussions in 
sects.~\ref{experim} and ~\ref{Analys}.

\begin{figure*}
  \begin{center}
     \vspace{+2mm}
    \includegraphics[angle=+00.0, width=0.63\textwidth]{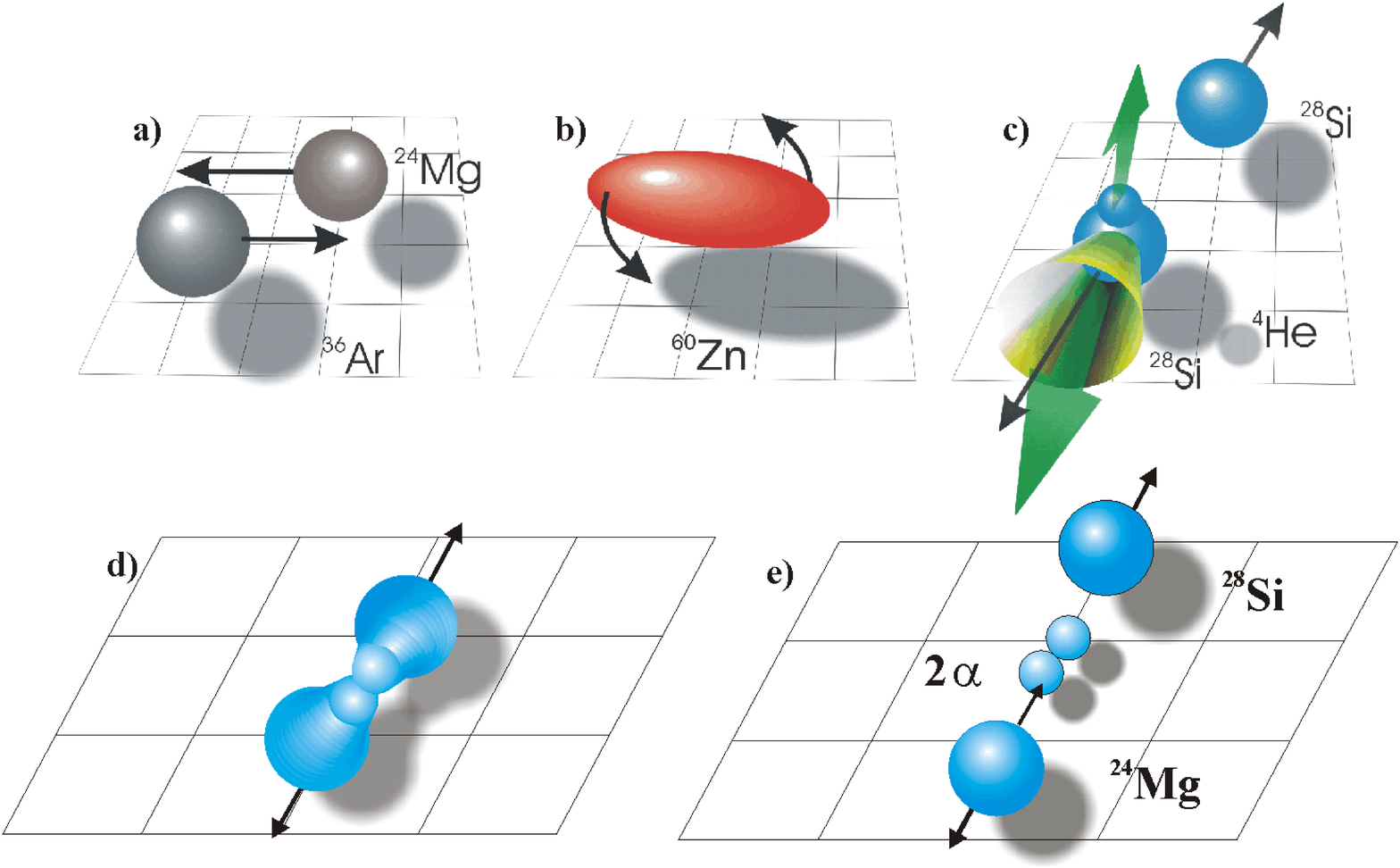}
         \vspace{-3mm}
    \caption{  (Color online)
              The binary and the ternary fission processes from possible 
              hyper-deformed states in the $^{60}$Zn compound nucleus
              with $\alpha$-cluster emission from the 
              binary fragments (c), or with the formation of
               $\alpha$-clusters in the neck (d,e), respectively. }
    \label{fig:TernaryFiss}
  \end{center}
\end{figure*}

The missing charges can be evaporated (statistical decay) from the excited
fragments in  binary fission which gives broad distributions for the 
in-plane and out-of-plane correlations.
If the missing charges are left with some smaller residual momentum 
between the two heavy 
fragments, we will refer to this process as ternary fission, see Fig.~\ref{fig:TernaryFiss}. 
The latter will give 
sharp correlations perpendicular to the reaction plane (coplanarity). 
The reaction plane is spanned by the beam axis and the vectors 
of the  two fragments in Det3 and Det4, with charges $Z_3,Z_4$.
For the following discussion the coplanarity
condition is of importance. We define  coplanarity with the relation for the 
out-of-plane angles,  ($\phi_3-\phi_4$)=180$^{\circ}$. 
The out-of-plane correlations, $N(\phi_3,\phi_4)$, of the
selected events are uniquely determined with the BRS-detector system over a very
wide angular range in the reaction plane ($\theta$).
These fragment yields $N(\phi_3,\phi_4)$ are very important
for the later discussion. 

The target consisted of  100 $\mu$g/cm$^2$  of  $^{24}$Mg enriched to 
99.9$\%$ on a layer of  20 $\mu$g/cm$^2$ of $^{12}$C.
The compound nuclei $^{60}$Zn are formed 
at an excitation energy of E$_{ex}$ = 88 MeV at a 
 bombarding energy of E$_{lab}$($^{36}$Ar) = 195 MeV. The experiment  performed at 
 the HMI Berlin and the BRS had rather good vacuum conditions. 
 Binary coincidences of the two  BRS-detectors  were registered with at least one
 $\gamma$-ray in coincidence in the OSIRIS-array. 
 The data have already been partially published in~\cite{thummerer98,zhereb06,zhereb07}.

\subsection{Experimental results}
\label{ArMgexp}
Charge separation has been obtained  in the BICs. The
reaction channels are defined using corresponding gates
on the BP-E distributions
for the different charges - thus  a large variety of  (Z$_{3}$+Z$_{4}$) combinations
are observed 
with their corresponding correlations in angles,  $\phi$ and  $\theta$. 
The individual binary 
and non-binary reaction channels can be clearly defined in this way.
The total kinetic energy (TKE) is the sum of the energies
of the registered fragments.
In order to judge the efficiency of the channel selection we  show the
inclusive yields in Fig.~\ref{fig:inclAr} with the two choices of 
the primary gate either in Det3 or Det4. The differences between the two distributions
 are due to the uncertainties in  the limits of the gates at the lower 
energies. For different charge-(mass)-asymmetries there are also  differences in the 
coincident  angular acceptance  in the $\theta$-ranges in the laboratory system.
 The fall-off of the intensities for large 
asymmetries is thus due to the loss in efficiency of registration 
(e.g. Figs.~\ref{fig:thetaAr} and ~\ref{fig:thet3_4ArMgodd}
in sect.~\ref{sec:inpl}).

A narrow distribution with  ($\phi_3-\phi_4$) = 180$^{\circ}$ uniquely defines 
the missing momentum in the reaction plane spanned by the 
beam direction and the 
vectors of the two heavy fragments. For $\Delta$Z = 4, 6 these narrow out-of-plane 
distributions 
define coplanar (or collinear) ternary fission with a small   out-of-plane momentum. 
As will be discussed later the decay of the elongated 
 CN at high angular momentum favours a collinearity of the fragments. 

For the determination of the kinetic energy of the fragments a
method for the energy calibration has been developed.
The general idea of the method is a transformation of the residual energy of the fragments in the 
Bragg Ionization Chamber (BIC) into their initial energy by using an analysis
 based on the calculation of the Bragg-curves.
The fragments are not fully stopped in the Bragg-chamber, and for charges smaller 16  ascending
and descending branches are obtained.
The new aspect of this procedure is the calculation of the mentioned branches of the 
Bragg-curves, to get the full energy information.
The calibration procedure includes several steps.

\begin{figure}[htbp]
  \begin{center}
    \vspace*{-8mm}
      \hspace*{-6mm} 
     \includegraphics[width=0.53\textwidth]{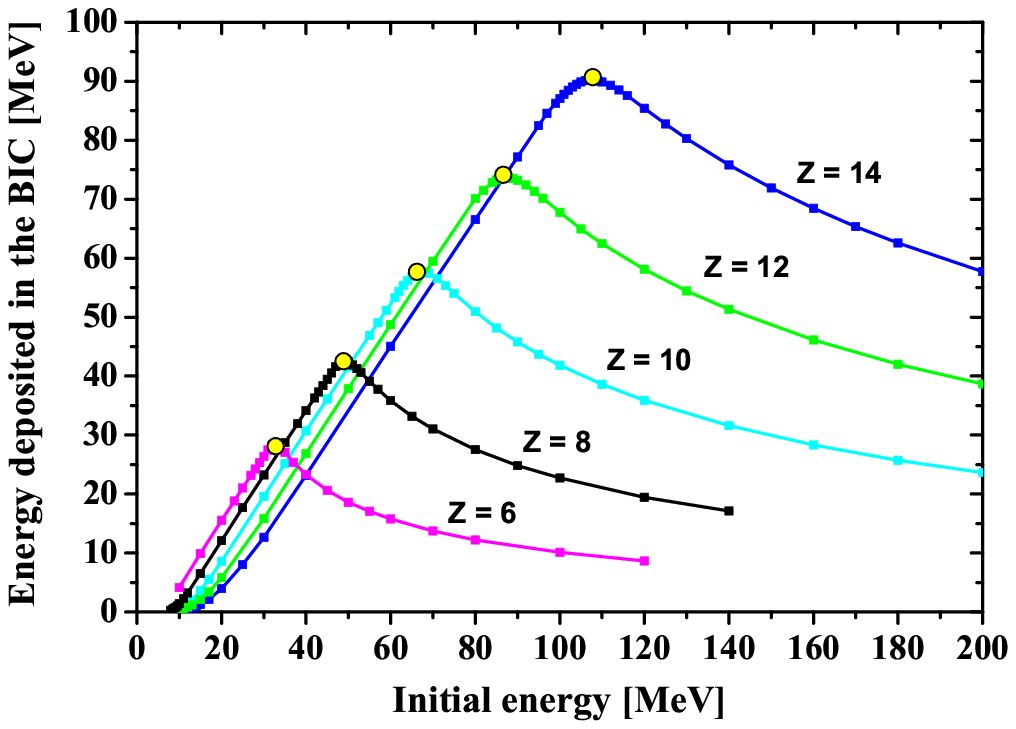}
    \vspace*{-10mm}
    \caption{ (Color online)
       The energy deposited by the  fragments in the BIC as a function of their 
       initial energy and the determination of the punch-through points (circles).}
    \vspace*{-6mm}
    \label{fig:ELoss}
  \end{center}
\end{figure}

1. Energy loss calculations for each charge were made with the identification
 of the punch-through point.
By using the code EVER (``Energieverlust Rechnungen'') the energy, which was lost by the
 fragments on the way from the target to BIC, has been calculated. 
Then the energy of the fragments deposited in the Bragg chamber as a function of the 
initial energy of the fragments was plotted for each charge (see Fig.~\ref{fig:ELoss}).

\begin{figure}[htbp]
  \begin{center}
     \includegraphics[width=0.45\textwidth]{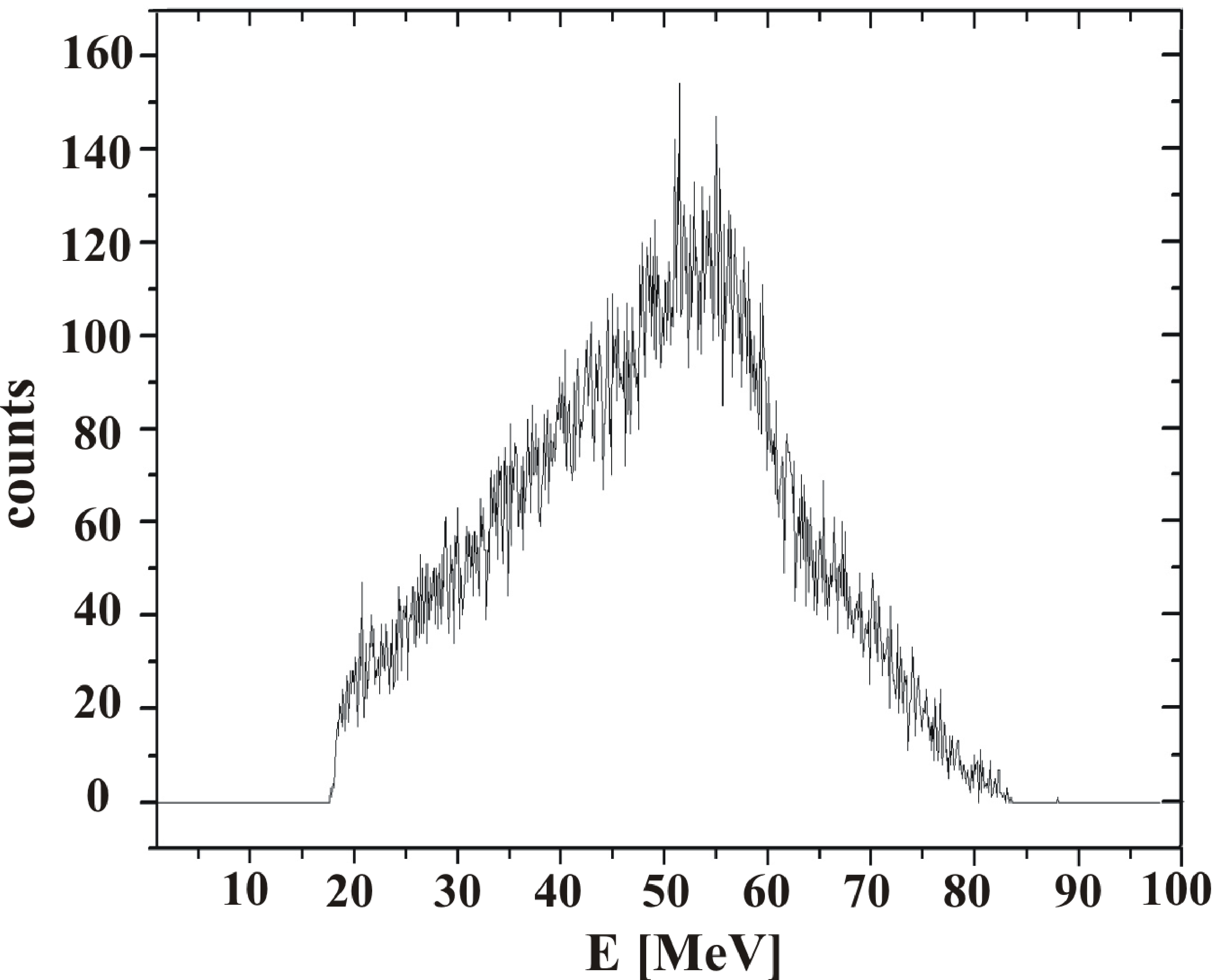}
    \vspace{-2mm}
    \caption{Energy spectrum of the $^{16}$O-nuclei registered in Det3, obtained with the 
        calibration procedure.}
   \label{fig:EnergO}
  \end{center}
\end{figure}

2. Search for the punch-through points in our experimental results (use 
of the Bragg Peak-Energy distributions).
Setting gates on the  back-bend region in the Bragg Peak (BP) - Energy (E)
 distributions 
we get the energy spectra for different charges 
 which contain the experimental punch-through point. 
Division of the experimental Bragg-Peak-Energy distributions 
into two branches: ascending and descending. 
The gates for the charges in the BP-E distributions were set in  accordance
 with the cuts for two different branches
(ascending and descending) in the Range-Bragg Peak distributions 
(for more detail see~\cite{zhereb07}).
A typical energy spectrum after this calibration procedure 
is shown in Fig.~\ref{fig:EnergO}. In order to estimate the uncertainty,
 we have compared the experimental 
maximum values of TKE for some combinations of  $(Z_3,Z_4)$
 with kinematical calculations 
and find that the experimental absolute values are typically 6-7 MeV to low. 
This effect is due to the energy loss calculations with 
differing values of the energy losses in the gas (which depend on temperature)
 and in the foils. The relative values 
have a much smaller uncertainty of less than 1 MeV. These considerations 
 are also relevant for the discussion in sect.~\ref{kinemat}.

\begin{figure*}[htbp]
\begin{center}
\vspace{1mm}
 \includegraphics[width=0.70\textwidth]{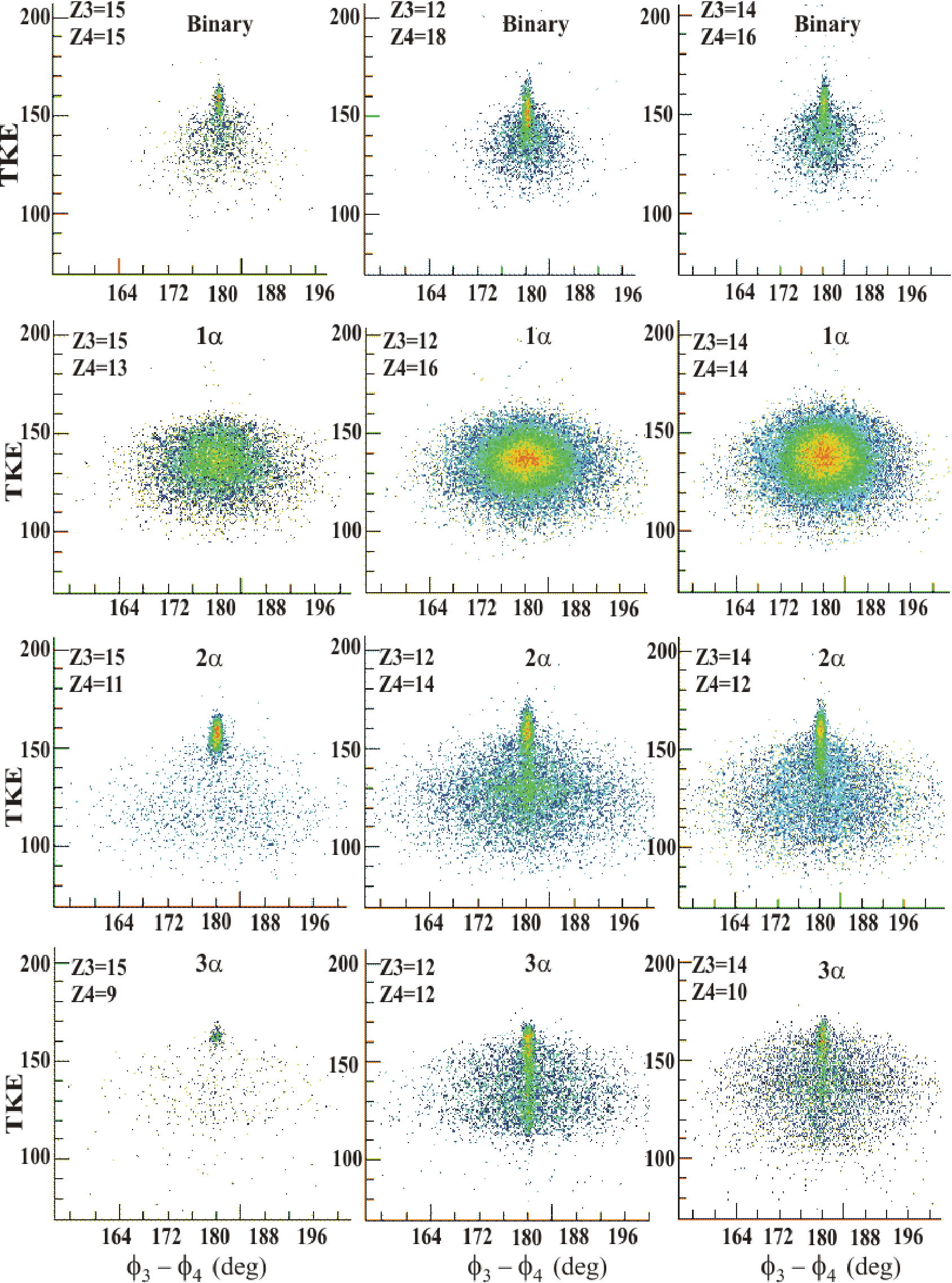}
\caption{ (Color online)
     Two-dimensional plots of the total kinetic energies (TKE in MeV, it is 
             not in an absolute scale) as function of the out-of-plane
             angles $(\phi_3-\phi_4)$ in degrees, for two coincident fragments 
             for which  the sum of the charges $Z3$ and $Z4$ is $even$, the
              missing number of 
              $\alpha$-particles is also indicated.
             }   
    \label{fig:TKEtwodimev}
\end{center}
\end{figure*}
\begin{figure*}[htbp]
\begin{center}
 \includegraphics[width=0.65\textwidth]{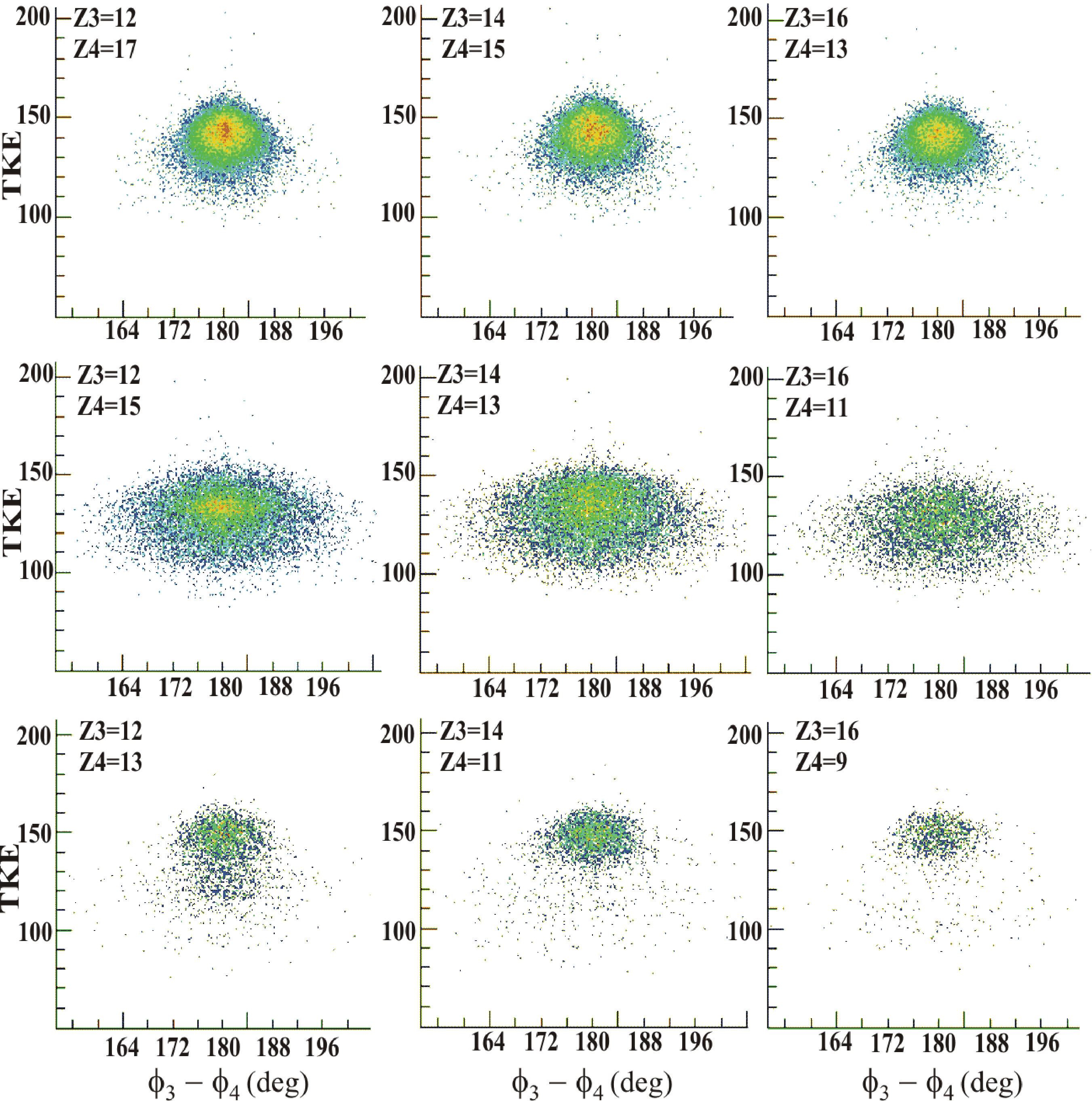}
      \caption{ (Color online)
             Two-dimensional plots (as in Fig.~\protect\ref{fig:TKEtwodimev})
            of the total kinetic energies (TKE in MeV, it is not in absolute 
             scale)  as  function of 
             $(\phi_3-\phi_4)$ (in degrees), for coincident fragments with 
             the sum of the charges
             $(Z3+Z4)$ = $odd$, as indicated.
             The missing numbers of 
             $\alpha$-particles and protons are from top: (-1p), (-1p,-1$\alpha$) 
             and (-1p,-2$\alpha$) 
             .}
    \label{fig:TKEtwodimodd}
\vspace{-3mm}
\end{center}
\end{figure*}

Using the present method of the energy calibration 
the total Kinetic energies (TKE, the sum  E$_3$+E$_4$) of the 
fragments have been obtained and 
two-dimensional TKE - $(\phi_3-\phi_4)$ distributions have been built 
(see Figs.~\ref{fig:TKEtwodimev} and ~\ref{fig:TKEtwodimodd}). The TKE values 
appear not in  an absolute scale due
to some remaining problems with the energy calibration of the BIC.

\begin{figure*}[htbp]
  \begin{center}
    \includegraphics[angle=+00.0,width=0.65\textwidth]{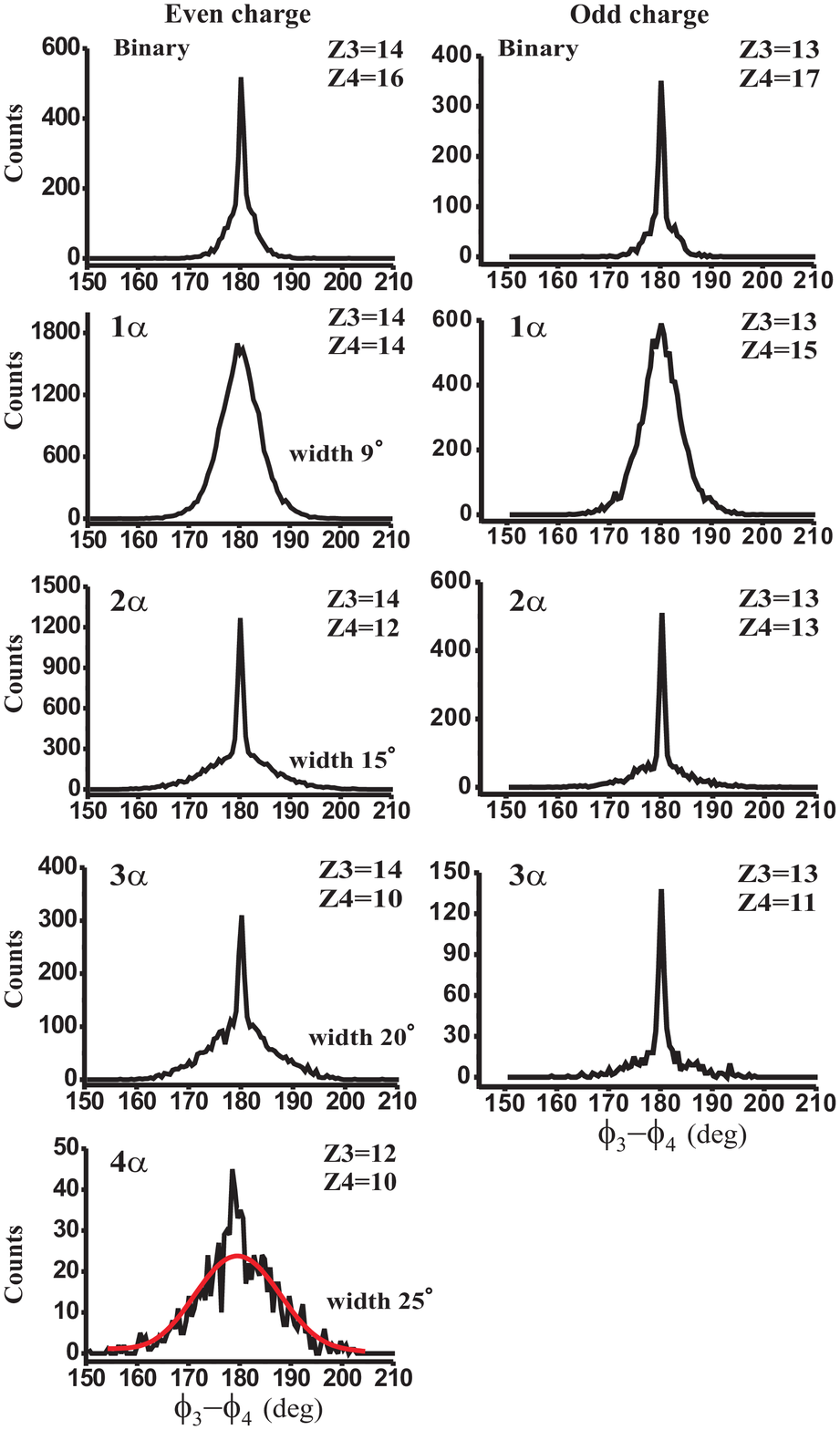}
    \caption{ (Color online)
             Yields, $N(Z_3,Z_4)$, of coincident fragments with charges
             indicated by $Z3$+$Z4$, as a function of the out-of-plane
             angles $(\phi_3-\phi_4)$ (in degrees).
             The width (FWHM) of  the
              angular correlations for the different fission channels are shown.}
    \label{fig:phiAr}
\end{center}
\end{figure*}

The TKE is shown 
as function of ($\phi_3-\phi_4$), for a pair of coincident charges $Z_3,Z_4$
with different combinations for the condition ($Z_3+Z_4$ = even or odd).
The same coincident fragment yields, $N(Z_3,Z_4)$, are 
shown in Fig.~\ref{fig:phiAr} as a projection on ($\phi_3-\phi_4$).
The distributions are fitted by Gaussians and the FWHM are given in the figure.

In the first row of Figs.~\ref{fig:TKEtwodimev} and ~\ref{fig:phiAr} we have the purely binary 
decay processes ($\Delta Z=0$) with  exit 
channels corresponding to different asymmetry 
of the charges. For purely binary ($\Delta Z=0$) events
narrow distributions around $(\phi_3-\phi_4)=180^o$ are indeed observed, a small
wider contribution results from the evaporation of  neutrons. 
No narrow distribution is observed for $\Delta Z=1, 2$. These events are
originally binary fission with an excitation energy in either fragment 
sufficiently high for a proton or one $\alpha$-particle to be emitted.
The distributions become wider with increasing number of missing  
$\alpha$-particles, the systematics of the widths are shown in Fig.~\ref{fig:phiAr}.
We can easily identify the broad yields as binary 
fission with increasing excitation energy in each primary fragment and the 
evaporation of 1 - 4 $\alpha$-particles, respectively. 
The Q-values allow the decay with $\Delta$Z = 8, see Fig.\ref{fig:phiAr},
however, these yields are very small. For asymmetric mass splits the coincident events 
move out of the angular acceptance in ($\theta_3,\theta_4)$. An example of these
correlations is given in Fig.~\ref{fig:thet3_4ArMgodd}.

\begin{figure*}[htbp]
  \begin{center}
      \includegraphics[width=0.95\textwidth]{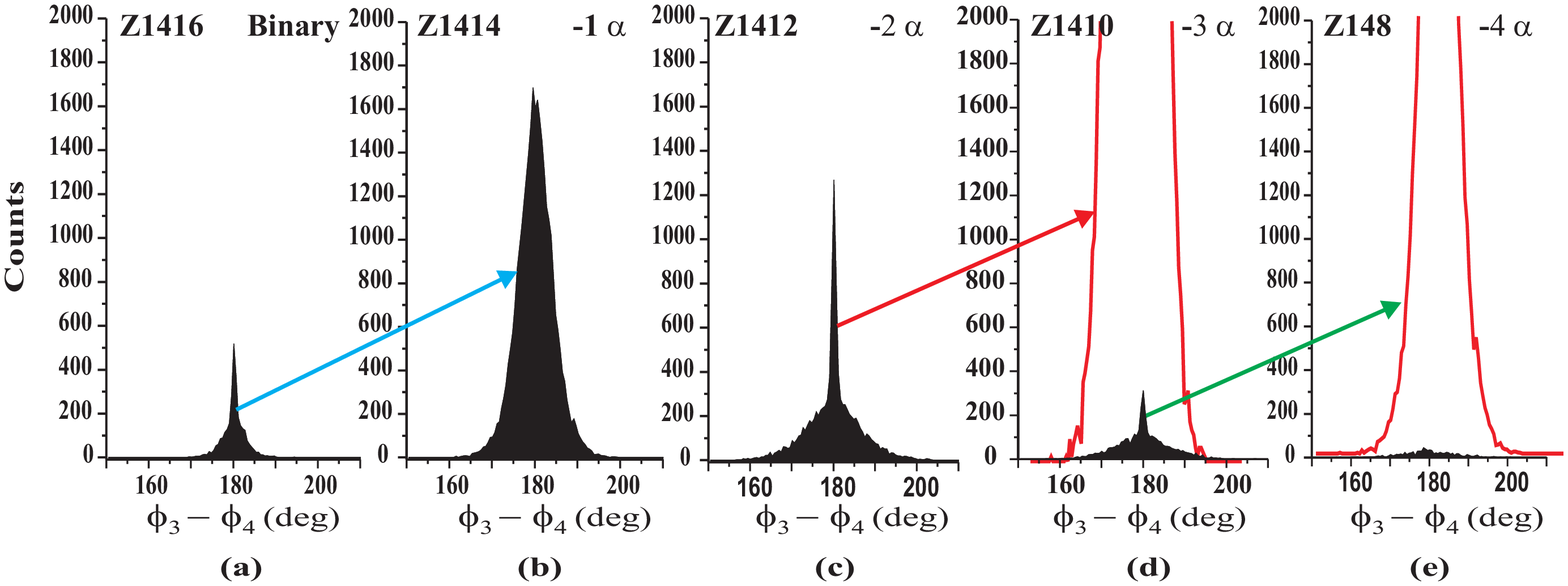}
    \caption{ (Color online)
             Out-of-plane angular ($(\phi_3-\phi_4)$, in degrees)  distributions
              for fission with definit values  of 
              $Z_3$+$Z_4$, with calculations showing 
              the expectation for the decays
               with -1$\alpha$ loss for $^{60}$Zn, and the assumption 
                of a narrow peak for $^{52}$Fe from  $^{16}$O in the target 
                and $^{48}$Cr from  $^{12}$C, respectively.
              The respective wide components for the determination of the
              contributions from contaminants in the target are shown 
              (see text for more details).   
             }
    \label{fig:contaminant}
  \end{center}
\end{figure*}

\begin{figure*}[htbp]
  \begin{center}
     \vspace{-4mm}
    \includegraphics[angle=+00.0,width=0.90\textwidth]{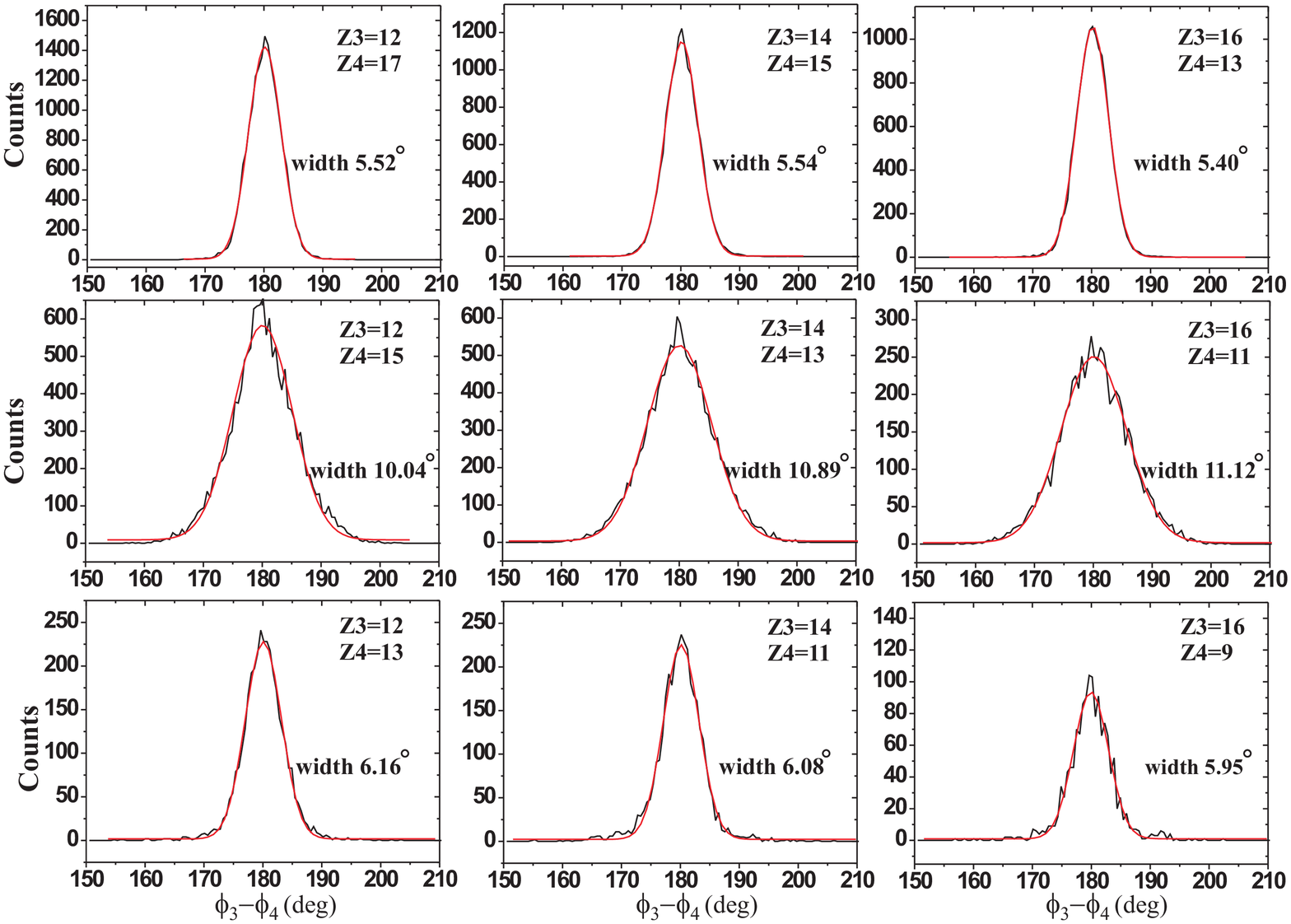}
    \caption{ (Color online)
             Out-of-plane  angular ($(\phi_3-\phi_4)$, in degrees) distributions  
             for coincident fragments with charges
             ($Z_3$+$Z_4$=odd), indicated as ($Z3$,$Z4$). 
              Top row - the process with 1p emission, middle row -
              with 1p+1$\alpha$ emission,
              bottom row - the process with 1p+2$\alpha$ emission.
             The Gaussian fit functions and the widths (FWHM) of the
              angular correlations are indicated.}
    \label{fig:phi1}
\end{center}
\end{figure*}

In the second row ($\Delta$Z$=2$) we observe the highest yield,
with a wide out-of-plane spectrum. For this particular case 
a fit by two different Gaussian  distributions (the first with the same 
width as for $\Delta$Z$=0$)
  gives a clear indication that a small narrow component 
exists in these channels (similar to the narrow parts observed for the larger $\Delta$Z). 
The contribution of the narrow parts
with a width as in the binary case is here about 2$\%$, the experimental ``narrow'' part
from neutron evaporation is about 9-10$\%$ of the total yield.
 
In the other rows with the coincidences for ($\Delta Z=2,4,6,8$), 
the ($\phi_3,\phi_4$)-correlations show an 
increasing width with larger charge losses,
the  broad component being located at lower TKE. This corresponds to the  expectation  
of a sequential uncorrelated statistical emission of
 several charged particles from the excited fragments. However,
for $\Delta Z=4$, which corresponds to two missing
$\alpha$-particles (third row in Figs.~\ref{fig:TKEtwodimev} and~\ref{fig:phiAr}) 
a strong  narrow component 
as in the binary cases is observed together with the broad
component. The pattern of narrow
correlations continues to appear for three  missing
$\alpha$-particles, where again two components, a narrow peak (as in the case
of a binary event) and a wider distribution,
  Fig.~\ref{fig:TKEtwodimev} and Fig.~\ref{fig:phiAr}, fourth row)
are observed. For the case of four missing  
$\alpha$-particles the Q-values are extremely negative, 
but again both components are visible (Fig.~\ref{fig:phiAr}).\\

We also show the yields,  TKE$-$$(\phi_3-\phi_4)$, for cases 
with an $odd$ missing charge ($Z3+Z4$ = $odd$), see 
Figs.~\ref{fig:TKEtwodimodd} and ~\ref{fig:phi1}. 
These correspond to an additional loss of a proton.  
The yields are lower due to the more negative Q-values. We note similar systematics, 
the -1p yields show clearly the effect of the statistical emission,  the yields 
with (-1$\alpha$,-1p) are again wider. However, only a small  
component appears in the channels with the loss of  (2$\alpha$s+1proton),
which have quite unfavourable Q-values for the  $^{24}$Mg-target. The main yield 
 at the higher TKE in these channels (of the bottom row in fig.~\ref{fig:TKEtwodimodd}) 
will originate from  reactions on $^{16}$O in the target, 
where it corresponds to the (-1p)-channel
with a much more favourable  Q-value (see Q-values in Fig.~\ref{fig:Qvalue_60Zn}).
This yield in comparison with 
those of the first row allows the determination of the $^{16}$O in the target, 
discussed in detail in sect.~\ref{Analys}.

\section{Analysis of the data}
\label{Analys}
\subsection{Out-of-plane correlations and target contaminants}
\label{outofpl}
The out-of-plane correlations allow a direct distinction between binary and coplanar ternary decays.
For the narrow component we have to discuss the contribution from contaminants in 
the target. This is a very important question, because the  effect observed in the 
out-of-plane distributions for (-2$\alpha$, $\Delta Z = 4$ and -3$\alpha$, $\Delta Z = 6$)  could
 be explained as a binary decay of a compound system formed with 
$^{16}$O or $^{12}$C instead of  $^{24}$Mg. 
In the present case it will be the $^{36}$Ar+$^{16}$O reaction (CN = $^{52}$Fe)
for the $\Delta$Z=4 channel, and the $^{36}$Ar+$^{12}$C reaction (CN = $^{48}$Cr) 
for the $\Delta$Z=6 channel.
 Therefore the probability of the reactions with $^{16}$O and 
$^{12}$C nuclei has been estimated. We have several  points, which show,
that the contributions due to $^{16}$O or $^{12}$C in the target are  only 15-20$\%$. \\
1. Estimations of  the absolute $^{16}$O-target thickness.  
For the target thickness determination of the $^{16}$O component, 
we assume that the 
 differential fission cross section  for the narrow part, binary for $^{16}$O 
 (with missing charge $\Delta Z=4$), to be the same as the  differential 
fission cross section for the pure binary decay  with $\Delta$Z=0
for CN=$^{60}$Zn. If  the narrow peak in the 
out-of-plane distribution for the $\Delta$Z=4 case arises only due to reactions 
on  $^{16}$O, with the same differential 
cross-section for fission  for both cases, we need two
 times more $^{16}$O than is possible for a completely oxidised $^{24}$Mg-target 
(only a 10$\%$ oxidation is expected). The vacuum in the BRS-system was below 10$^{-6}$ mbar 
in the present experiment.
 
2. Analysis of the ratios of the differential cross sections for 
the binary  ($\Delta$Z=0) to the   $\Delta$Z=2 charge
combinations.
From the out-of-plane distributions (see Fig.~\ref{fig:contaminant},(a))
 we have calculated the ratio   R(1$\alpha$/0$\alpha$), in the c.m.-system of the differential 
cross sections of the binary fission with 1$\alpha$ emission 
($^{36}$Ar+$^{24}$Mg $\rightarrow$ $^{60}$Zn
$\rightarrow$  $^{28}$Si+$^{28}$Si +1$\alpha$) to the differential cross-section 
 of the pure binary process
($^{36}$Ar+$^{24}$Mg $\rightarrow$ $^{28}$Si+$^{32}$S). The same value  is assumed 
for this ratio for the  decay of the CN= $^{52}$Fe (the  $^{16}$O-target). 
We find that the yield in the broad distribution 
 $\Delta$Z=2,  
is  8 times stronger (in Fig.~\ref{fig:contaminant},(b)), 
than the yield of the pure binary ($\Delta$Z=0) decay,  
 R(1$\alpha$/0$\alpha$,$^{60}$Zn)=8. 
Applying this value to the narrow part in $\Delta$Z=4 with the contaminant, for 
$^{36}$Ar+$^{16}$O $\rightarrow$ $^{28}$Si+$^{24}$Mg, the broad component in $\Delta$Z=6 
from a 1$\alpha$ emission (the reaction
$^{36}$Ar+$^{16}$O $\rightarrow$ $^{28}$Si+$^{20}$Ne+1$\alpha$) should be 8
times bigger than this narrow part, (Fig.~\ref{fig:contaminant},(d)). 
However, the experimental results show a rather small broad component, it 
is a factor 10 too small 
to originate from $^{16}$O. A similar argument applies for $^{12}$C in the target 
(see Fig.~\ref{fig:contaminant},(d, e)),
 the broad distribution for $\Delta$Z=8  is to small to be due to the
$^{36}$Ar + $^{12}$C $\rightarrow$ $^{48}$Cr  $\rightarrow$ 
$^{28}$Si+$^{16}$O+1$\alpha$) reaction.
A maximum contribution of  15 $\%$ to the narrow component in $\Delta$Z=6 is estimated.

3. The systematics of the widths of the out-of-plane distributions 
for the binary decays with sequential $\alpha$-emission have been analysed.
The broad out-of-plane correlations are expected 
 to have an  increasing width for increasing missing charges.
 This is indeed fulfilled for the distributions with missing charges 
of $\Delta$Z=2, $\Delta$Z=4,~6 and
 $\Delta$Z=8, the width increases from 9$^\circ$ to 25$^\circ$, respectively,  
as shown in Fig.~\ref{fig:phiAr}) and  Fig.~\ref{fig:contaminant}.
This fact shows that the observed broad components are due to the $^{36}$Ar+$^{24}$Mg
 $\rightarrow$ $^{28}$Si+$^{24}$Mg +X$\alpha$ reaction, 
where the X$\alpha$-particles are emitted from excited fragments.

\begin{figure*}
  \begin{center}
    \includegraphics[width=0.69\textwidth]{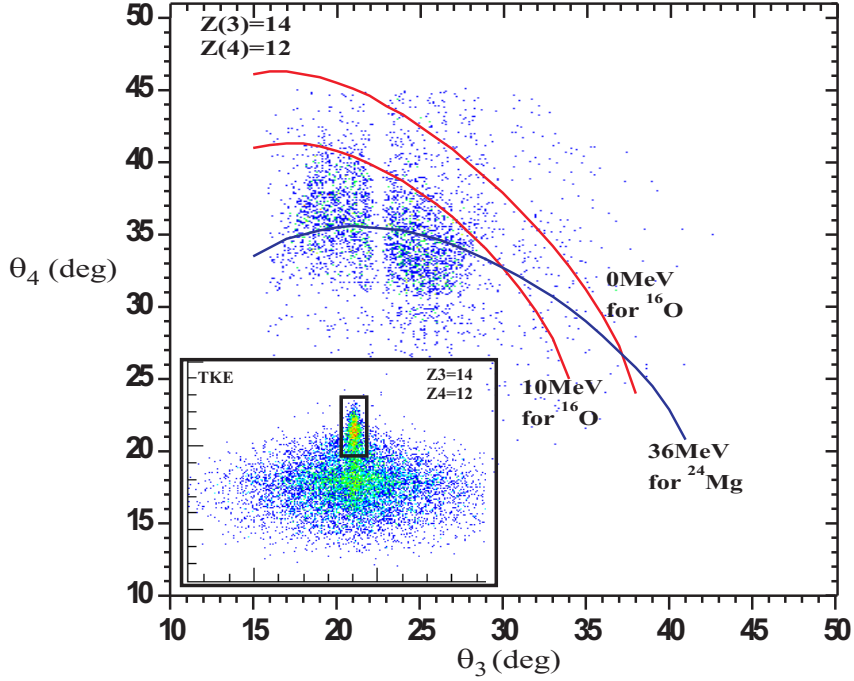}
           \vspace{-2mm}
    \caption{  (Color online)
             In-plane $\theta_{4}$ vs. $\theta_{3}$ (in degrees) correlations , 
             of coincident fragments for the channels gated on the  
             narrow (ternary) components for definit values  
              $Z_3$+$Z_4$ for  the
              fission defined by $Z_3$+$Z_4$+2$\alpha$. 
              The kinematical calculations are shown for the Mg-target 
              and for the O-target (channel $Z_3$+$Z_4$+0$\alpha$ 
              representing a binary channel) with the 
              effective Q-values as indicated and  as discussed in the text. 
              The angular range in the centre of mass frame 
               is between $\theta_{c.m.}$=65$^\circ$ to 120$^\circ$.}
    \label{fig:thet3_4Ar1}
  \end{center}
  \vspace*{-3mm}
\end{figure*}

4. Analysis of the in-plane angular distributions ($\theta_{3}$-$\theta_{4}$).
The analysis of these two-dimensional correlations  has been done
 together with kinematical calculations for the reactions, where the targets are
 either $^{16}$O or $^{24}$Mg. Using a gate applied in the total kinetic energy
 (TKE)-out-of-plane correlation, we have taken only those events,
 which produced the narrow part. 
Then, the kinematical curves for the same fragments (Z$_{3}$=14, Z$_{4}$=12), formed either in 
the $^{36}$Ar + $^{16}$O or
$^{36}$Ar + $^{24}$Mg reactions have been calculated. We use an appropriate excitation 
energy E$_{ex}$ of the
 fragments to allow  $\alpha$-particle evaporation. In the case of 
 $^{36}$Ar + $^{16}$O reaction
with (Q$_{eff}$ =~10 MeV), with Q$_{eff}$ = Q$_{0}$ + E$_{ex}$, and 
 E$_{ex}$ must be chosen to be below the decay threshold, but allowing  mutual excitation.
The latter kinematical curve lies in 
the border region of the experimental events in the two-dimensional angular 
distribution for the narrow part with $\Delta$Z=4, 
see Fig.~\ref{fig:thet3_4Ar1}. In the region
 with a further increase of the excitation energy (Q$_{eff}$ =~20 MeV), fragments will emit 
 $\alpha$-particles, and the experimental yield of the $\theta_{3}$ vs. $\theta_{4}$
 distribution  can not anymore correspond to the  exit channel
 (Z$_{3}$=14, Z$_{4}$=12) with a $^{16}$O-target. 
 Thus the main yield originates from the $^{24}$Mg-target, for the latter  
 we do not have restrictions in the excitation energy, i.e. from 
 a reaction channel with the excitation energy sufficiently high in 
each fragment to allow evaporation of two $\alpha$-particles (Q$_{eff}$ $\approx$ 36MeV).
For this reaction the kinematical curve is in good agreement with
 the region where the main intensity of the 
 registered events are located.

5. As a further point we analyse the out-of-plane correlations
for the cases with  $odd$ total charges, like 
the  ($^{36}$Ar+$^{24}$Mg$\rightarrow$$Z_3+Z_4$+1p+x$\alpha$)-channels, 
see Fig.~\ref{fig:phi1}. 
Here also an increase of the width as in 
Fig. \ref{fig:phiAr} with increasing charge loss is expected. However, the 
channel (-1p-2$\alpha$) has the same width as 
the (-1p)-channel (see Fig.~\ref{fig:phi1} bottom and top rows).
Due to this observation it can not
originate from the reaction on the  $^{24}$Mg-target, which actually has a very  
unfavorable Q-value and only a small cross section is expected. We can conclude that  
these events originate  from  the $^{16}$O-contaminant, because the  corresponding 
Q-values for the ($^{36}$Ar+$^{16}$O$\rightarrow$$Z_3+Z_4$+1p)-reaction
on $^{16}$O are only (-10 -14) MeV and the reaction with 1p-evaporation 
can proceed without hindrance (see Fig.~\ref{fig:Qvalue_60Zn}).
 With this observation and making 
the assumption that the binary fission channels of $^{60}$Zn and of $^{52}$Fe 
with the evaporation of one proton to be equal, we can determine a thickness of 
16.8 $\mu$g/$cm^2$ (17$\%$) of $^{16}$O in the target.
This result is in very good agreement with our previous
determinations. From these estimations we can also calculate the differential cross sections
of the different exit channels of the reaction
$^{36}$Ar+$^{16}$O$\rightarrow$$Z_3+Z_4$+1p, which are shown in Fig.~\ref{fig:CrSecOdd}.\\ 
6. The higher kinetic energy of the narrow component. As discussed later 
(see sect.~\ref{kinemat}) for the ternary decay the
third fragments are assumed to be created in the neck, and will have rather 
low kinetic energy in the centre of mass system.
 We have performed calculations of the 
corresponding three-body decay. These cases imply for example that
 the missing particle, an $^{8}$Be nucleus or $^{12}$C,  is 
emitted ``backwards'' in a fast sequential decay from one of
the moving heavier fragments.

 For the statistical evaporation with all directions allowed,
a ``circular'' broad distribution is obtained  in the TKE-$(\phi_3-\phi_4)$ plot. 
The collinear geometry 
gives us the upper energy limit of the TKE of these  distributions.
The resulting values of the TKE are in the range of 27-35 MeV higher than
 the centre of  the broad component of the energy distributions (see Fig.~\ref{fig:shift}).

To summarise these observations we conclude that the narrow yields originate
dominantly from fission  of $^{60}$Zn ($^{24}$Mg target),
approx. 15-17$\%$ of the narrow distributions can originate from target
contaminants.

\subsection{In-plane correlations, angular distributions}
\label{sec:inpl}
\begin{figure}
  \begin{center}
     \vspace*{-6mm}
       \hspace*{-8mm}
    \includegraphics[width=0.57\textwidth]{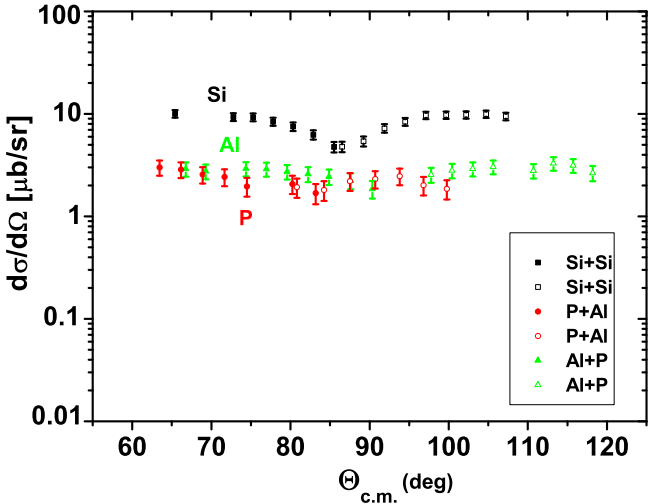}
      \vspace*{-8mm}
    \caption{ 
            (Color online) Differential cross sections for coincident fragments  for some  
             channels (defined by 
             $Z_3$+$Z_4$+1$\alpha$) representing the binary  fission  
               (broad)  components.
             The angular ranges  in the centre-of-mass frame
              are between  $\theta_{c.m.}$=65$^\circ$ to 120$^\circ$.}
    \label{fig:angdistr}
  \end{center}
 \vspace*{-4mm}
\end{figure}

For the case of  $^{36}$Ar + $^{24}$Mg we show some angular 
distributions of the binary channels 
 in Fig.~\ref{fig:angdistr}.
These differential cross sections can be obtained by 
taking either Det3 or Det4 as the primary gate, the corresponding 
symbols in the figure show these choices. In the range of the unrestricted efficiency 
of the coincident measurements, the angular distributions are
rather flat and symmetric around 90$^\circ$.
 This result points to the fact that
the binary decay is a fission process from an equilibrated  compound nucleus.
For the fusion-fission mechanism  the angular distributions of the final
fragments should be isotropic in the centre-of-mass frame, with a rise towards small angles.
 A deviation from symmetry (and isotropy) will
characterise processes with the formation of a di-nucleus system and a fixed direction for the
emission of the fragments would arise in such processes.

\begin{figure*}[htbp]
  \begin{center}
     \vspace{3mm}
    \includegraphics[angle=+00.0,width=0.65\textwidth]{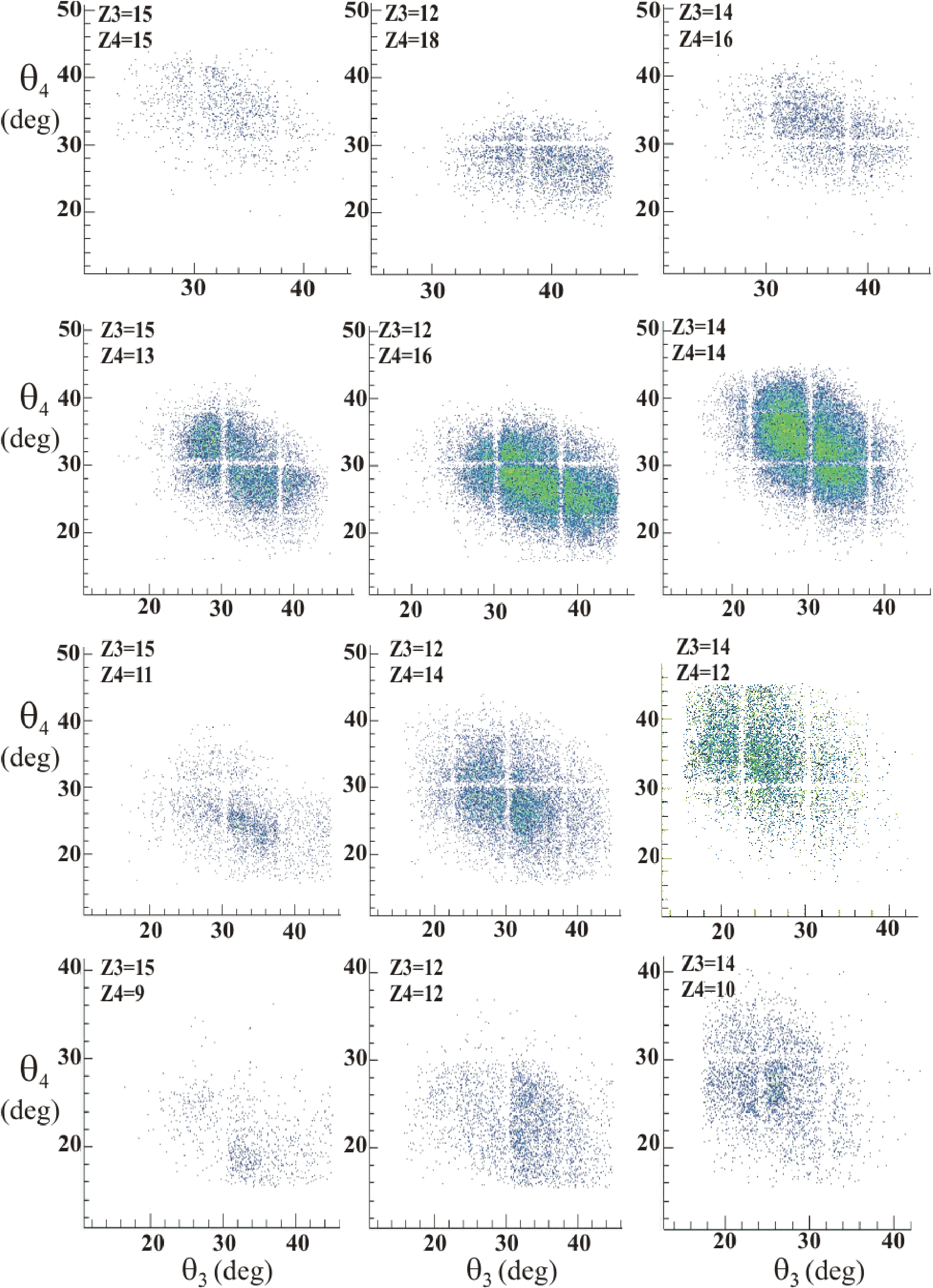}
    \caption{  (Color online)
             Correlations of the in-plane
             angles  $(\theta3,\theta4)$ (in degrees) for 
             coincident  fragments with charges
             indicated as $Z_3$,$Z_4$, for 
             $^{36}$Ar + $^{24}$Mg at $E_{lab}=195~\textrm{MeV}$.
             The empty regions are due to gaps in the electronically defined read-out. 
             }
    \label{fig:thetaAr}
  \end{center}
\end{figure*}

\begin{figure*}
  \begin{center}
    \includegraphics[width=0.68\textwidth]{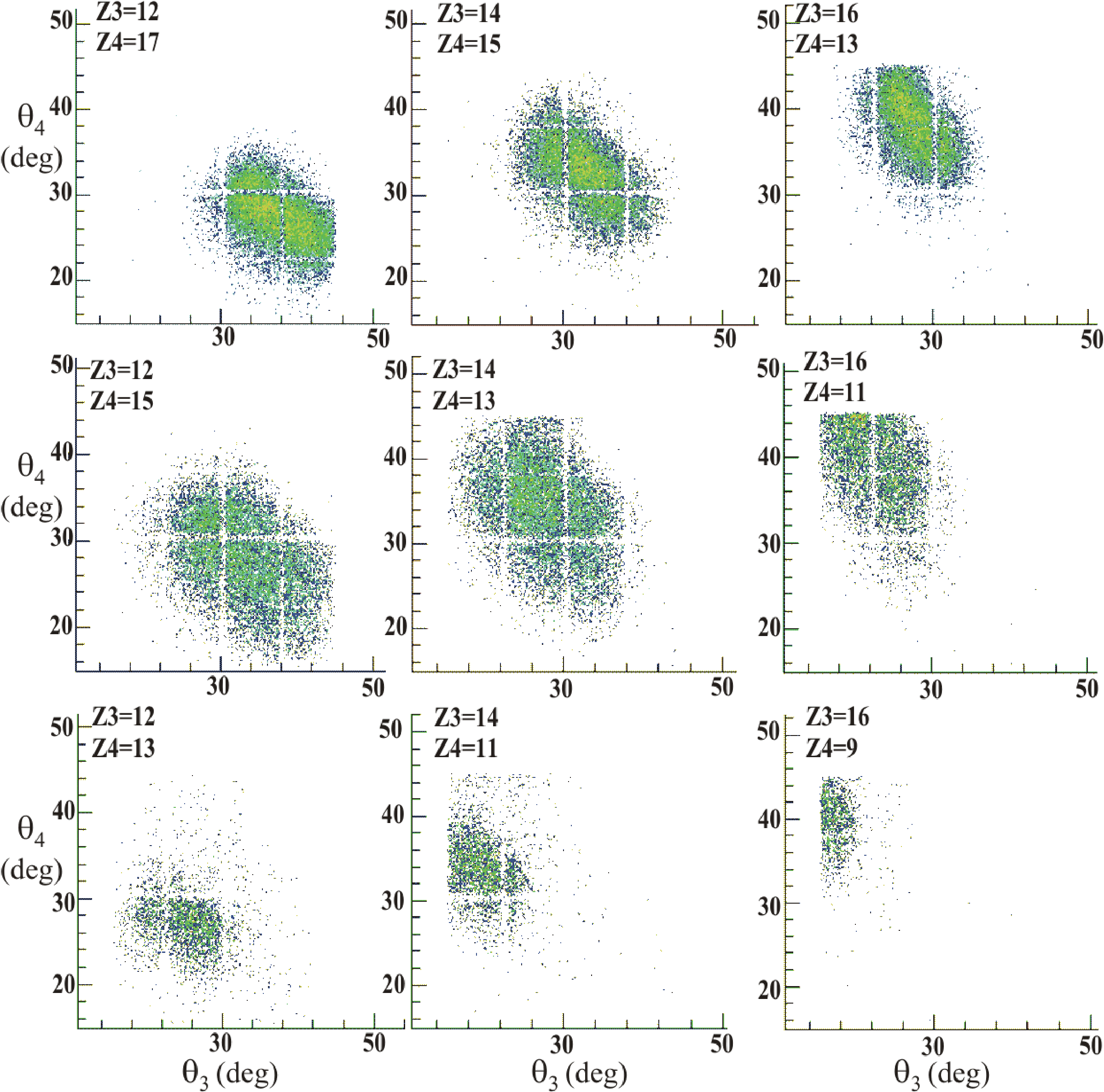}
    \caption{  (Color online)
                In-plane experimental angle-angle correlations of fragments  for
              channels of binary  coincident 
              fission events  defined by 
              ($Z_3$+$Z_4$+1proton+x$\alpha$). 
              The centre-of-mass   angular ranges 
             are between  $\theta_{c.m.}$ = 65$^\circ$ to 120$^\circ$.           
              }
    \label{fig:thet3_4ArMgodd}
  \end{center}
\end{figure*}

Another feature, which can  be used to characterize the reactions is the 
folding angles ($\theta_{3}$, $\theta_{4}$)$-$correlations in the reaction plane  of
the two coincident fragments. 
The correlations in the  ($\theta_{3}$, $\theta_{4}$)-plane reflect the different 
geometric coincidence efficiencies of the BRS. For symmetric mass splits 
and corresponding (negative) Q-values the experimental choice of 
the angular ranges of the BRS covers perfectly the regions of interest. For more 
asymmetric mass splits this region eventually moves out of the experimental 
($\theta_{3}$, $\theta_{4}$)-range, this is illustrated by the angular correlations 
in the Figs.~\ref{fig:thetaAr} and~\ref{fig:thet3_4ArMgodd}.

Of further interest is the comparison with kinematic calculations of the 
observed binary coincidences, already mentioned before. For the  
(Z$_{3}$=14(Si)+Z$_{4}$=12(Mg))-channel 
shown in Fig.~\ref{fig:thet3_4Ar1}, the calculated  kinematic lines are shown 
together with the kinematic lines for the 
reactions on  $^{16}$O, which is a  binary channel.
 In order to have the possibility of two $\alpha$-particles 
evaporated the corresponding excitation energy in the fragments 
should be added to the ground state Q-values.  
Very negative 
Q-values for the reaction on $^{24}$Mg have to be assumed ($Q_{eff}=Q_{0}-25$ MeV).
 The emission of the  $\alpha$-particles 
corresponds to a broad region in the in-plane and out-of-plane angular distributions.
As discussed before a separation of events beyond the kinematical line for $^{16}$O
is possible, and the remaining events 
(the dominant part) are due to the $^{24}$Mg-target. 
 
Summarising  all these conditions we are able to extract the 
differential cross sections for binary (broad components) and
ternary (narrow) fission yields. 
\begin{figure*}
  \begin{center}
     \vspace{-4mm}
        \includegraphics[width=0.95\textwidth]{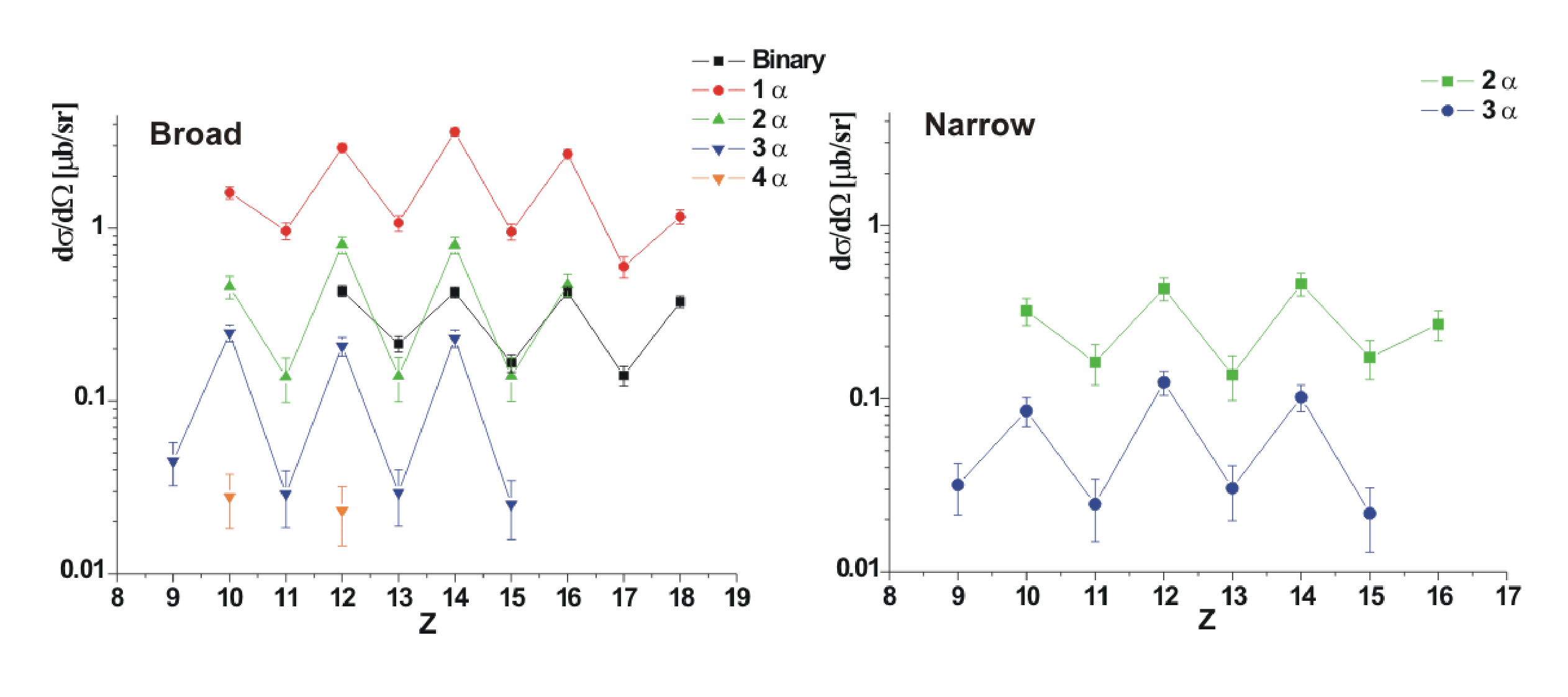}
         \vspace{-4mm}
          \caption{  (Color online)
              Differential cross sections in the reaction
             $^{36}$Ar + $^{24}$Mg for channels defined by 
              ($Z_3$+$Z_4$+x$\alpha$)  
              of the broad (binary) and narrow (ternary) components of the 
             coincident fission fragments. They  represent the average over the
             angular range between $\theta_{c.m.}$ = 65$^\circ$ to 120$^\circ$ in 
             the centre-of-mass system.
              The fission channels 
              with missing 1$\alpha$, 2$\alpha$, 3$\alpha$ and 4$\alpha$-particles are  
              shown.
              }
    \label{fig:yieldsAr}
  \end{center}
\end{figure*}

\subsection{Differential cross sections}
\label{xsection}
\begin{figure*}
  \begin{center}
       \includegraphics[width=0.72\textwidth]{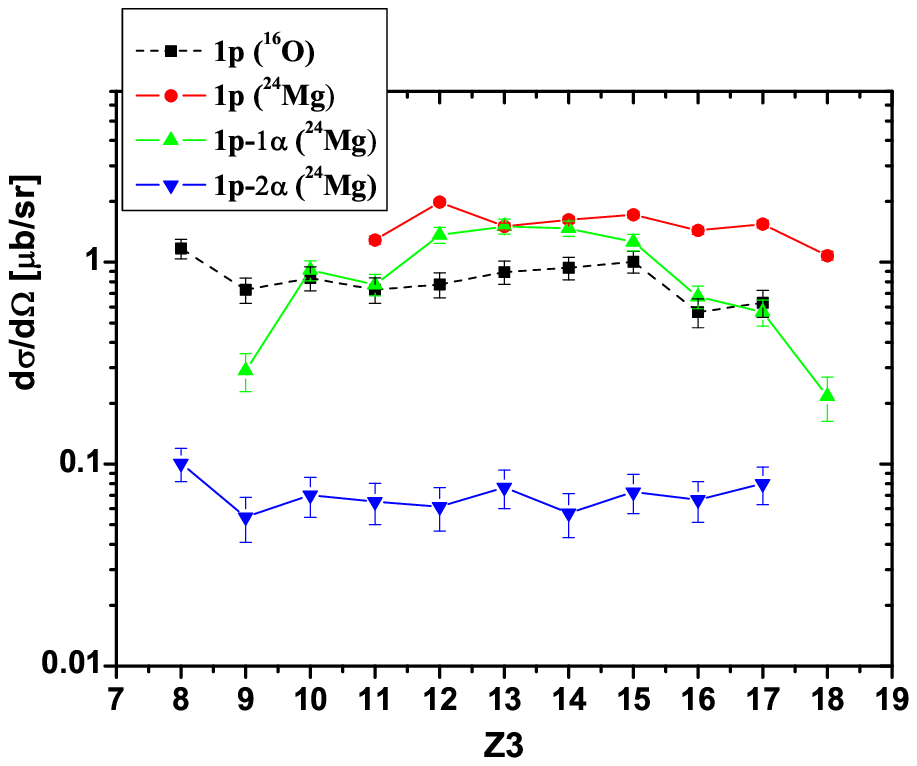}
         \vspace{-9mm}
              \caption{  (Color online)
              Differential cross sections for the reaction
             $^{36}$Ar + $^{24}$Mg as in Fig.~\ref{fig:yieldsAr} 
             for channels defined by $Z_3$+$Z_4$+1p+x$\alpha$, 
            and the differential cross section in the reaction
             $^{36}$Ar + $^{16}$O for channels defined by 
              $Z_3$+$Z_4$+1p
               The fission channels 
             with missing -1p, -1p-1$\alpha$, -1p-2$\alpha$ for $^{24}$Mg
             target and -1p for $^{16}$O target are  
              shown.}
    \label{fig:CrSecOdd}
  \end{center}
\end{figure*}

As shown in the previous section, the angular distributions are rather flat, 
therefore we can give average values 
of  the differential cross sections for  the different exclusive binary and ternary channels 
in the range of the observed centre-of-mass angles, $\theta_{c.m.}$.
 These ranges are approximately  from 65$^{\circ}$ to 120$^{\circ}$ with a variation 
of 10$^{\circ}$, depending on the values of $(Z_3,Z_4)$. 
We have plotted the yields (differential cross sections)
of the broad distributions for the exclusive binary fission
channels in Fig.~\ref{fig:yieldsAr}, left part and for the ternary fission, right part.
The yields have been analysed as a function of  charge asymmetry for the different
values of missing charges ($\Delta$Z, $\Delta$Z=even).  In Fig.~\ref{fig:yieldsAr} on the X-axis
the charges of the fragments registered in  Det3 are plotted.
The fragments in the Det4 should be: ${Z_4} $= $Z_{CN}$-${Z_3}$-$\Delta$Z. Similarly 
 the yields for $\Delta$Z=$odd$, i.e. for the processes with emission of 1p, 1p+1$\alpha$,
1p+2$\alpha$, are given in  Fig.~\ref{fig:CrSecOdd}. 
There the exclusive differential cross-sections for the $^{24}$Mg-target are 
shown together with the differential cross-sections for the elements 
which are produced from the reaction on $^{16}$O in the 
target. We repeat our result, that 17${\%}$  contributions of reactions on 
$^{16}$O (as contaminant in the target) are contained in the values shown in Fig.~\ref{fig:yieldsAr}. 

\begin{figure*}
  \begin{center}
     \vspace{-8mm}
     \includegraphics[width=0.75\textwidth]{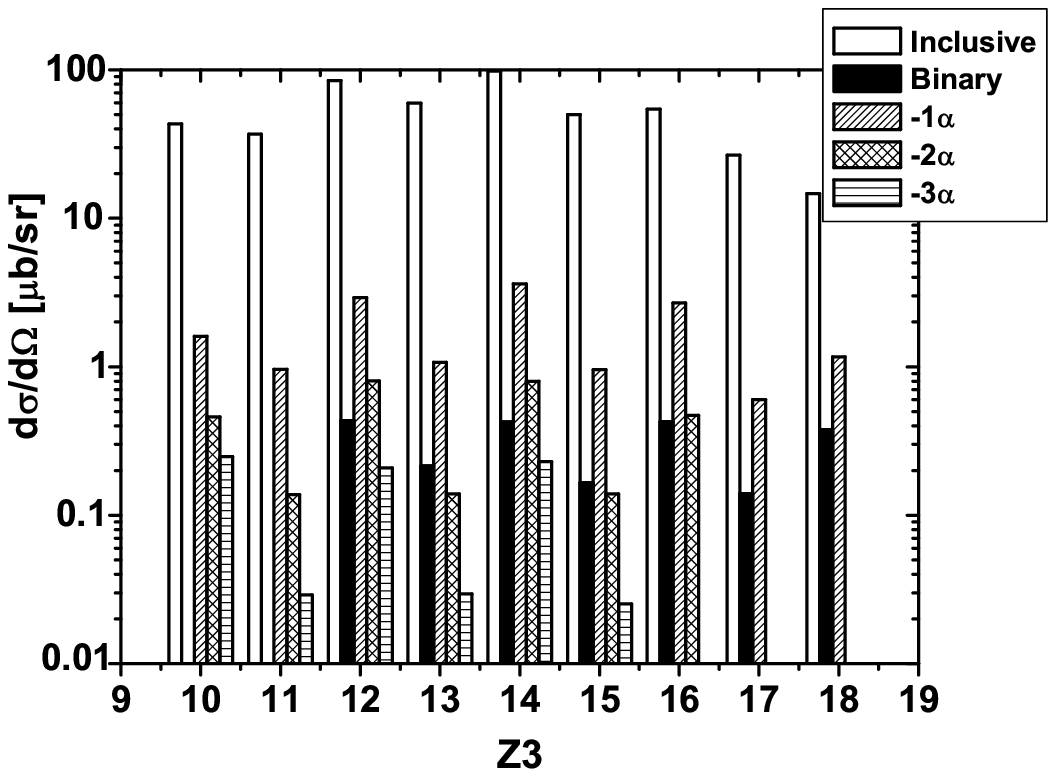}
      \vspace{-4mm}
         \caption{ The inclusive (open histogram) and exclusive 
              (solid and shaded histograms) diferential cross-sections
                  for fission (in the angular range 
                 $\theta_{c.m.}$ = 65$^\circ$ to 120$^\circ$) in 
                  the $^{36}$Ar+$^{24}$Mg reaction.}
    \label{fig:Histogramm}
  \end{center}
\end{figure*}

The yields as a function of the charge-asymmetry should be symmetric around the 
value $(Z_3+Z_4)/2$  for either combination  of $(Z_3,Z_4)$. 
 This symmetry is well respected except for the 
larger asymmetries, where deviations by  a factor 1.5 - 2 appear due to reduced
 kinematical efficiencies and differences in  setting gates at the low E-values.
The result shows that 
 the efficiencies and the gates chosen 
in the data reduction  correspond to the same energy ranges. We also note that the absolute values
of the cross sections of the  present work are almost the same (factors 1.5 - 2) as
those obtained in the independent experiment with the BRS-Euroball-setup for the 
fission decay of  $^{56}$Ni in the  $^{32}$S + $^{24}$Mg reaction (the values 
of the angular momentum and excitation energy are similar)~\cite{Efimov07,voe08}.
The  inclusive and exclusive cross-sections are summarized in a 
 histogram in Fig.~\ref{fig:Histogramm}.

\section{Interpretation, ternary fission}
\label{inter}
\subsection{Reaction mechanisms}
\label{reacmech}
Before we explore the interpretation of the data as compound nucleus deacy, we shortly discuss
a few other reaction mechanisms for the binary and ternary cluster decay.
Most important in the present work are the out-of-plane correlations
for events which cover a large angular range in $\phi$ and $\theta$.
This was only possible due to the large solid angle of the BRS and its 
high spacial resolution in the X- and Y-coordinates.
The broad component in the out-of-plane distributions with $\Delta$Z=2-8 can be understood as
a statistical $\alpha$-emission process where 1-4 $\alpha$-particles are emitted from the
fully accelerated fragments. In the momentum space the fragments after emission have a kinematical
angular cone of the scattering. The mechanism of this process has been 
 shown schematically in Fig.~\ref{fig:TernaryFiss}.
The narrow components around ($\phi_3-\phi_4$) = 180$^\circ$ can 
originate from different fission mechanisms assuming either: 

i) a fission process after a fast emission, for example of one or two (and even more)
$\alpha$-particles, a process which will slightly disturb the observed correlation of the two
subsequently emitted fission fragments (pre-fission emission). \\
ii) a mechanism where, e.g. four charge units (the case of two missing $\alpha$-
particles)  must be emitted,  correlated in-plane by a process involving two primary
heavy ejectiles, with their angular momenta strongly aligned perpendicular to
the reaction plane. \\
iii) Ternary fission with the ``emission'' of the missing charges from the neck,
the missing mass (X$\alpha$-particles) is emitted backward (in a sequential process, see below)
in the center-of-mass frame. This process produces a narrow ($\phi_3,\phi_4$)-correlation as in a
standard binary decay, because the neck-particles carry small  momentum perpendicular  
to the reaction plane. The emission angle ($\phi_3-\phi_4)$ remains 180$^{\circ}$,
this process can define coplanar or collinear decays (see Fig.~\ref{fig:TernaryFiss}).\\

For the first case we have to evaluate the second chance  fission probability. 
One can use the values of the evaporation cross sections from the detailed
analysis of the CN decay which have been done 
by Sanders et al.~\cite{sanders99}. The compound nucleus decays are mainly by particle
evaporation, giving rise to dominantly residual masses of A=48-50, (emission of
one or two $\alpha$-particles and some nucleons). This result is consistent
with the systematics of Morgenstern~\cite{morgenstern}, that the average energy
needed to emit one nucleon is  16.4 MeV and for one $\alpha$-particle is 23.4 MeV. 
Similarly  in the emission of
one fast (pre-equilibrium) $\alpha$-particle this amount of energy or more will 
be removed. For binary fission to occur after the emission of a first particle, 
the fission probability is decreased drastically, because of the reduced excitation
energy, and no second chance fission can be expected in  such a  
reaction. Indeed no significant contribution from a narrow peak in the
($\phi_3,\phi_4$)-correlations are observed for the fragment-fragment
coincidences with one missing $\alpha$-particle ($\Delta$Z=2) or 
one missing charge ($\Delta$Z=1).   The arguments against the role
of pre-scission $\alpha$-particles are even more severe in the cases  of two
pre-fission $\alpha$-particles and we can rule out this process. 

For the second scenario involving a correlated emission of the $\alpha$-particles  
from the two fragments we can argue as follows.
The fact that the narrow correlations appear as strong for $\Delta$Z=4 and 6
makes it rather unlikely that such a very special correlation 
persists through all decays. The correlation created in the 
first binary decay must be complete and has to persist  
in the sequence of secondary emissions from the fragments for 
an odd and even  number of emitted $\alpha$-particles. 

For the third scenario we assume a  ternary fission 
process with the third clustered fragment in the neck (which will  consist of 
$\alpha$-particles). If the two heavier 
fragments are emitted in a coplanar (or collinear) geometry with the third fragment, 
a correlation  as sharp
as in the case of a binary fission process is expected. This will happen naturally for
the highest angular momenta, and only in these the condition for the height of the saddle points 
allows a competition between ternary and binary decays as shown in the next subsect.\ref{kinemat}.
Only this scenario can give a consistent image of the observed 
decay characteristics. 

\subsection{Kinematics of binary and ternary fission}
\label{kinemat}
For the discussion of the ternary fission mechanism we will
consider the formation and the decay of the 
compound nucleus at the highest angular momenta, as also discussed in ref.~\cite{zhukov81}. 
In this work a three cluster system with an interaction potential between the clusters as 
obtained from the experimental data, is considered, and it is shown
that for the small angular momenta a triangular configuration is favored. For
higher angular  momenta the decay system has  a 
stretched configuration caused by centrifugal forces.
For the ternary cluster decay in  a three-body configuration 
at high angular momentum, the energy is minimized
if the lighter fragment is placed between of the two heavier fragments.
The collinear configuration
has the highest moment of inertia and the lowest energy for the saddle point. This fact
 produces the narrow component in the ($\phi$) correlations at $(\phi_3-\phi_4)$=180$^{\circ}$. 
The distinction between the broad and narrow components can be viewed as due
 to two different time scales
 of the ternary cluster decay. For the broad component the secondary 
cluster emission occurs 
after the two heavier fragments have been accelerated by the decay energy 
(Coulomb and centrifugal potentials). The narrow component is obtained if in the 
fission process two neck ruptures occur in a short time sequence.

\begin{figure*}
  \begin{center}
    \includegraphics[width=0.82\textwidth]{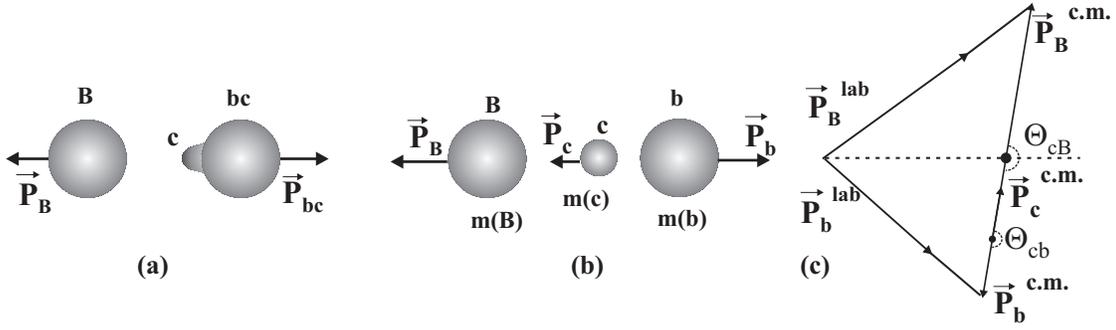}
    \caption{Sequential fission decay: Ternary fission as a consecutive fast 
          two-step process.
             The first step  - (a),  a binary 
             decay of the hyper deformed compound system in two fragments $B + (b + c)$.
               The second step - (b) - the decay of  $(b + c)$,
              cluster  $C$ emission from one of the fragments, (c) the
              vectors in the laboratory  and the centre of mass systems. 
              The configuration is assumed to be collinear at high angular momentum.   }
    \label{fig:sequ}
  \end{center} 
\end{figure*}

As a model calculation for ternary fission we will  discuss the decay of the CN 
 $^{60}$Zn with 2$\alpha$ or 3$\alpha$-clusters formed in the neck.
This decay proceeds through two phases: first - decay of the compound system into
two highly excited fragments. We actually may assume a population of the resonant 
cluster-states in these fragments. 
The second step - the sequential emission of the ternary fragment ($\alpha$-clusters, cluster,c), 
from one of the fragments. 
We write schematically the reaction as:
$ A + a \rightarrow B + (b+c)$, see Fig.~\ref{fig:sequ}, 
with the fragments $B$ and $b$  finally observed. The collinear decay 
 mechanism  can be modeled with
 the $\alpha$-clusters  being emitted "backwards" in the second step in 
a very short time sequence, comparable to the fission time.
 In this case 
 we will assume that the primary angle between fragments  $b$ and $B$ remains 180$^{\circ}$ 
in the center-of-mass frame, as shown in Fig.~\ref{fig:sequ}, part (b). In the alternative 
decay with the statistical emission of two or three 
$\alpha$-clusters from highly excited fragments,
 the primary emission angle is changed into an emission cone (see Fig.~\ref{fig:TernaryFiss}),
due to the recoil momentum induced by the  $\alpha$-clusters. The $average$ value of the
angle would not change, as well as the average momentum, giving an avarage value of TKE corresponding
to the total Q-value ($Q_{eff}$). We note that for the statistical decay in addition to
the ground state Q-value an excitation energy in the primary fragment sufficiently high above 
the Coulomb barrier for the emission is needed. In the following the kinematics of 
the ternary decay is discussed.

As an example we take the 2$\alpha$-emission process, and we perform our calculations
 in the centre-of-mass frame of the compound nucleus. We define $\vec{P_{B}}$, $\vec{P_{bc}}$,
 $\vec{P_{b}}$ and $\vec{P_{c}}$ as the momenta of the first primary fragment - B, 
of the second primary fragment - bc, of the final fragment b 
(after cluster emission) and of the ternary cluster - c, correspondingly.
 Analogously we define: $E_{B}$, $E_{bc}$, $E_{b}$, and $E_{c}$ - the kinetic
 energies in the exit channel. \\
For the first phase of the reaction, energy and momentum conservation gives  
 with $Q_{eff}$ = $Q_{0}$ + $E^{*}$: \\
$\vec{P_{B}}$ + $\vec{P_{bc}}$ = 0, or 
\begin{displaymath}
\left\{\begin{array}{ll}
E_{B} + E_{bc} = E_{c.m.} + Q_{eff} & \\
m(B)E_{B} = m(bc)E_{bc}.  &              \\
\end{array}
\right.
\end{displaymath}
From here we define the energy of the two fragments
\begin{equation}
E_{B} = \frac{m(bc)\cdot[E_{c.m.}+Q_{eff}]}{m(bc)+m(B)} \\
\end{equation}
  
\begin{equation}
E_{bc} = \frac{m(B)\cdot[E_{c.m.}+Q_{eff}]}{m(bc)+m(B)}.\\
\end{equation}  
In the second phase, when $\alpha$-clusters are emitted from the fragment $bc$ 
(see Fig.~\ref{fig:sequ}, parts b, c),
the three-body kinematics in the centre-of-mass frame 
 for the collinear cluster decay of the compound system $^{60}$Zn has to be calculated. 
Energy conservation for this reaction gives:
 $E_{B}$ + $E_{b}$ + $E_{c}$ = $E_{c.m.}$ + $Q_{eff}$. The conservation 
of the linear momentum gives:         
 \begin{equation}
\vec{P_{B}} + \vec{P_{b}} + \vec{P_{c}} = 0.\\
\end{equation} 

The kinetic energy of one of the fragment is then: 
\begin{multline}
E_{b}=\frac{(\vec{P_{B}} + \vec{P_{c}})^2}{2m(b)} = 
\frac{\vec{P_{B}}^2 + \vec{P_{c}}^2 + 2\vec{P_{B}}\vec{P_{c}}}{2m(b)}=\\ 
=\frac{m(B)}{m(b)}E_{B}+\frac{m(c)}{m(b)}E_{c}+\\
+\frac{2\sqrt{m(B)m(c)}}{m(b)}\sqrt{E_{B}E_{c}}\cos\theta_{B c},
\end{multline}
where $\theta_{B c}$ is the angle between $\vec{P_{B}}$ and $\vec{P_{c}}$,
 in our case $\cos\theta_{B c}$=1.
 The equation in velocity units is:
  \begin{multline}
E_{b}=\frac{m^2(B)}{2m(b)}\upsilon_B^2+\frac{m^2(c)}{2m(b)}
\upsilon_{c}^2+\frac{2m(B)m(c)}{m(b)}\upsilon_{c}\upsilon_B  
\end{multline}
The velocities $\upsilon_{B}$, $\upsilon_{c}$
 have been calculated. For the definition  of $\upsilon_{c}$ we consider the 
 decay of the primary fragment bc into fragment b and cluster c. Then the 
conservation of the linear momentum for this case is $\vec{P_{b}}$ = - $\vec{P_{c}}$  
For the energy we have : $E_{c}$ + $E_{b}$ = $E^{*}_{bc}$ where $E^{*}_{bc}$ = $E^{cm}_{bc}$ + Q,
 - the excitation energy of fragment bc. The energy of c is:  
\begin{displaymath}
E_{c} = \frac{m(b)\cdot[E^{*}_{bc}]}{m(b)+m(c)},\\
\end{displaymath}
 and from here we define $\upsilon_{c}$. 

\begin{figure*}
  \begin{center}
     \vspace{-5mm}
    \includegraphics[width=0.70\textwidth]{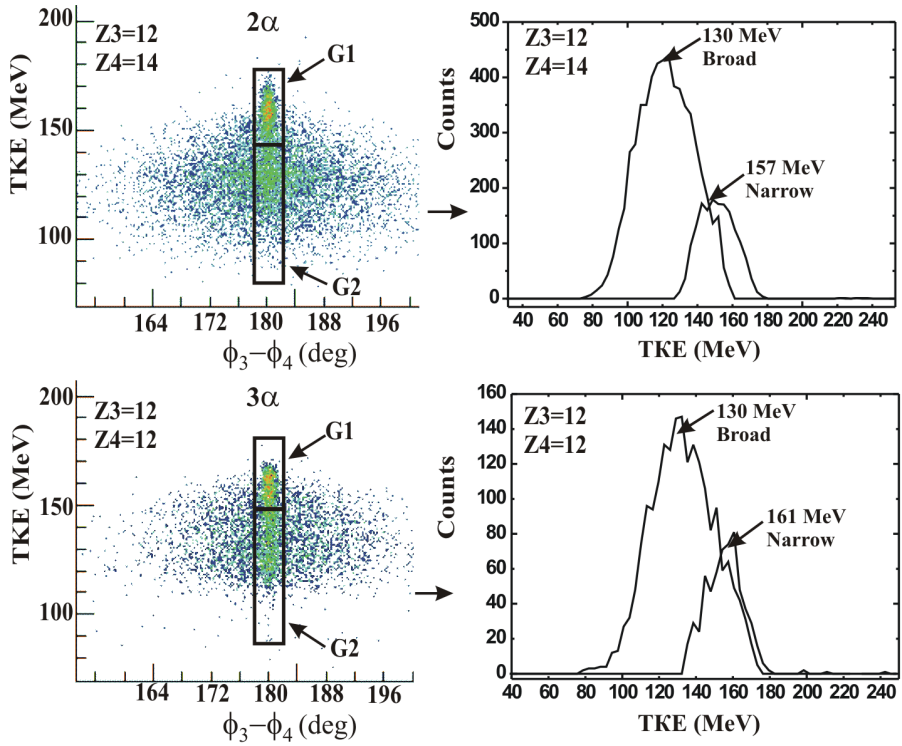}
     \vspace{-2mm}
    \caption{ (Color online) Two dimensional TKE distributions (left part) 
            vs. out-of-plane angles (in degrees) 
            $(\phi_3-\phi_4)$ 
           and the projection of the TKE with two separated: broad and narrow 
           components (by using gate 1 - G1), respectively (right part). 
           The gate 2 (G2) is for the checking procedure,
            see text for details.}
    \label{fig:shift}
  \end{center} 
\end{figure*}
The calculations have been carried out for the case $^{36}$Ar + $^{24}$Mg $\rightarrow$
 $^{60}$Zn{*}  $\rightarrow$ $^{24}$Mg + $^{36}$Ar{*} $\rightarrow$ 
($^{24}$Mg + $^{28}$Si + 2$\alpha$). 
From the calculations we find that in the above mentioned model for ternary fission as a 
sequential process, the TKE values are larger than compared to the average value 
of the TKE of the fragments produced in the binary decay processes with the random emission 
of $\alpha$-particles from the moving fragments. In fact,
 we expect that the ternary fission process produces energies of the fragments $b$, 
which are placed at the upper limit of the almost ``circular'' distributions of the 
TKE-($\phi_3,\phi_4$)$-$correlations (see Fig.~\ref{fig:TKEtwodimev}). 

For the narrow part 
in the out-of-plane correlations 
we can compare the result of the calculations with the experimental data. 
The energy in the experimental data was obtained in the lab. system,
 but the difference between the TKE of the ternary fission and of the energies after 
the $\alpha$-evaporation 
processes can be compared with calculations for the TKE in the centre-of-mass frame. 
Using gates G1 (see Fig.~\ref{fig:shift}, left part), we have separated 
the narrow and broad components in the experimental data and have observed, 
that the average difference of TKE between the narrow and the broad parts is 
(for c=2$\alpha$) approximately 27 MeV, see Fig.~\ref{fig:shift}, right part, top. 
This value agrees with the calculations where the difference is 30 MeV. The experimental 
relative values have an uncertainty of 1 MeV, as discussed in sect.~\ref{ArMgexp}.
In the same way calculations were performed for the ternary 3$\alpha$-cluster decay: 
$^{36}$Ar + $^{24}$Mg $\rightarrow$($^{24}$Mg + $^{24}$Mg + 3$\alpha$).
 We predict, that the TKE of the fragments 
 are 33 MeV higher than the average value in the binary decay with statistical 
3$\alpha$ emission from fragments. This is again in  good agreement with 
the experimental result.  
The experimental difference is 31 MeV, see Fig.~\ref{fig:shift}, right part, 
bottom. In order to check that this result does not depend strongly on the lower 
border of the gates we have opened the gate to the lower TKE values (see Fig.~\ref{fig:shift}, G2). 
The average difference of TKE between the whole narrow and the broad parts has remained the same (actually the counting rate around the peaks did not change, differences
smaller than 100 KeV) as it was before (with using G1).
Therefore, the dominant yield in the data
is concentrated in the chosen gates.
From these results we 
conclude that the collinear geometry describes
 the  the TKE-($\phi_3,\phi_4$)$-$correlations
in an appropriate way. The experimentally wide distributions in TKE actually suggest
that there is also a large component of ternary cluster decay with coplanar geometry
with the collinearity as the limiting case. This may be connected to a distribution of 
fission times as suggested in sect.\ref{statist}.

As a final statement we may consider that the final fragments  $B$ and $b$ 
in the case of a ternary cluster decay should have rather low excitation 
energy (i.e. they are ``cold'' fragments) 
 as compared to the binary case. In contrast for  the latter case,  as expected from
the statistical phase space considerations (see below),  
the highest possible  excitation energy 
(with the maximum of level densities) of the primary fragments is populated and a feeding 
 into a high level density after particle emission is expected also in the final fragments. 

\subsection{Statistical model fission decay}
\label{statist}
From  the present experimental results we conclude that the 
cluster decay of an equilibrated CN is observed.
One of the main  point in the interpretation of the data is
to explain the observed yield of the ``ternary'' fission channels
 relative to the binary decay. The basic Q-values for the ternary mass split, 
and the corresponding evaporation process are the same, however, for the 
evaporation to occur the fragment must be excited to a higher energy 
above the separation energy, namely the corresponding Coulomb barriers should 
be included. This makes the Q$_{eff}$ for the statistical evaporation 
10 MeV higher (or more) and in the statistical model an advantage for
 the direct ternary mass split may occur.

For  the interpretation of the data as a   fission process,
we have to consider the statistical  phase space 
for both binary and the coplanar ternary fission. 
This can be achieved 
by the Extended Hauser-Feshbach Method (EHFM)~\cite{matsuse97}
which has been developed for heavy-ion fusion-fission reactions in order to provide
a detailed analysis of all the possible decay channels by including the fusion-fission phase space.
The EHFM supposes that the fission probability is proportional to the available ``phase
space at the scission point''.
The narrow width and the high values of the TKE of 
 ternary events give us opportunity to make a model,
where the formation N$\alpha$-particles in the neck is assumed. We will
disregard the phase space of the particles in the neck, as 
well as their kinetic energy. The initial stages for the ternary and binary
 fission may be considered to be the same, the difference being in the time scales
 of the separation of the ternary clusters, which are emitted earlier in a coplanar
 or collinear configuration.
Energy conservation then gives: 
\begin{center}
{\bf $E_{CN}^*$ = $Q_{tern}$ + $(U_3 +U_4 - E_{3,4})$}
\end{center}
for a compound nucleus (CN) of an excitation energy {\bf $E^*_{CN}$}, with a ternary Q-value  
, $Q_{tern}$, 
the excitation energies of the fragments are given by {\bf $U_3$, $U_4$} with 
their relative kinetic energy as $E_{3,4}$. 

The  differential decay cross section for different  mass partitions 
$(i,j)$ will depend on  the product of the level densities $\rho_i(U_i,J_i)$,
 for $i=3,j=4$ of the fragments and the   
total inverse fusion cross section  $\sigma(E_{i,j},R_s,J)$ 
\begin{equation}
\frac{d\sigma(i,j)}{d\Omega dE_{k}(i,j)} = C{\rho}_{1}(U_i,J_i){\rho}_{2}(U_j,J_j)
\sigma(E_{3,4}),R_s,J)  \nonumber
\end{equation}
we use $J =  J_{3} + J_{4}$, for the spins of the fragments,  with total spin  $J$.
The radial variable $R_s$ ($R_s$=$R_3$+$R_4$+d, where d is a neck parameter) 
stands here for the 
shape of the saddle, and we  defined the ``inverse'' formation probability
$\sigma(E_{3,4},R_s,J)$ as in the usual fission theory, where it belongs to 
the corresponding fusion cross section. 
We have  the constraints for the excitation energies 
and the kinetic energy: ($U_3$+$U_4$ =  $E^*_{CN}$ - $Q_{tern,3,4} - E_{3,4})$.
Or expressed as an effective Q-value,  $Q_{eff,3,4}=Q_{tern,3,4}+U_3+U_4$.
In the ternary decay the two excitation energies of both  heavy fragments (3,4), which  
are registered in coincidence before further decay,  must have values  
$U_{3,4}$ below their $\alpha$-decay threshold.
 In our case for  even-even fragments with (N=Z) 
the thresholds for a evaporation process (including a Coulomb barrier) 
are in  the region of  (10-15 MeV).
 In the statistical 
model, see e.g. ref.~\cite{sanders99}, the  spin dependent level densities 
in the two fragments are determined by their excitation energies. 
 The effective Q-values are very important and determine the relative
 kinetic energy $E_{kin(i,j)}$=$E_{3,4}$ and thus the corresponding 
penetrabilities $T(E_{kin(i,j)},R_s,J)$. For these the potential energy surfaces 
for the two fragments must be considered.
The formation cross section  in the exit channel,
 $\sigma(E_{kin}(i,j),R_s,J)$ for a
 given $E^*_{CN}$ and $J$ contains the 
 different  penetrabilities  $T(E_{kin(i,j)},R_s,J)$ in the 
exit channels and we make the model assumption of an inverse fusion given by:
\begin{equation}
 \sigma(E_{CN}^*,J,Q_{tern,(i,j)}) = 
\int d E_{kin} \delta U d \Omega \frac{d\sigma}{d\Omega  \delta U dE_{kin}}
\end{equation}
with $\delta U$ = $U_3$-$U_4$, which 
defines the excitation energy differences for the two fragments.
The  value of $\sigma$ is thus strongly influenced  by the ternary Q-values 
of  the  channels $(i,j)$. For the determination of  
$T(E_{kin}(i,j),R_s,J)$  the potential energy surfaces 
of the fission path leading to the fission  saddle point have to be evaluated.

At this point a comment on the geometrical shape of the fission saddle and 
 the three-body configuration is needed. The three fragments could be placed 
at different relative orientations, however, it can easily be shown that for 
larger values of the total spin $J$, the collinear configuration which 
has the largest moment on inertia relative to all others, will have the
 lowest barrier. This feature has been calculated for some specific 
orientations of the three fragments
 by Wiebecke and Zhukov~\cite{zhukov81}, where it 
 is shown that the lowest barrier occurs for the collinear configuration.
Thus we may indeed expect the collinear ternary cluster decay and the competition
between binary and ternary fission becomes visible only at higher angular momenta.

For the total  energy balance which includes these potential energies 
and their dependence  of the rotational energy due to the deformed shape 
we introduce the free energy
 $(i,j = 3,4)$: 
\begin{equation}
 E_{free}(i,j) = E_{CN}^* +   Q_{eff,3,4} + 
V_{pot}^{eff}(J,Z_3,Z_4,R_s)
\end{equation}
The available  free energy $E_{free}(i,j)$ will be used 
to populate regions of maximum level density in the fragments,
it will determine the yield for a particular partition. Most important is 
 the total potential energy at
 the saddle point determined by  --  $V_{pot}(J,Z_3,Z_4,R_s)$. The latter  contains the 
rotational energy  $E_{rot}(J,3,4,R_s)$, which depends on  
the total spin $J$ and the moment of inertia $\Theta_{ff}(R_s)$ and finally on 
the shell corrections $\Delta_{sh}(R_s)$ expected for particular shapes 
at the  deformed  saddle point,
\begin{equation}
V_{pot}^{eff}(Z_3,Z_4,R_s) = E_{rot}(J,3,4,R_s)+
V_{pot}(Z_3,Z_4,R_s)+
\nonumber\\
\end{equation}
\begin{equation}
+\Delta_{sh}(R_s) 
\end{equation}
The shell corrections  $\Delta_{sh}(R_s)$ for hyper-deformed shapes
 can be read from the review of Ragnarson, Nilsson and Sheline~\cite{Rag81},
they are in the range of 3-8 MeV.

The available energy to populate states in the fragment can now be 
determined as in a binary configuration considering the potential energy
surfaces as function of deformation or fragment distances. To summarise,  the positions of
the fission barriers (and scission points) of different channels is
governed by their Q-values, by the liquid drop energies and by 
the rotational energies for a given $J$. These values are shown 
for some channels in Table.~\ref{tab:1}.
Using this approach the interpretation of the relative strength of the binary 
and ternary fission yields can be obtained. According to the statistical model 
the differences between ternary and binary mass split arise from:\\ 
a) different Q-values, and therefore different height of the
barriers, but also due to different values of $U_i,J_i$,\\
b) total  fission barrier heights due to
the different moments of inertia $\Theta_{ff}$ (see table \ref{tab:1})
and due\\
c) the shell corrections for large  deformations (with 2:1 and 3:1 axis ratios).
\begin{table}[htbp]
\caption{ Q-values, the inverse of the moments of inertia 
          (${\hbar}^2/2\Theta_{ff}$) and the barrier heights (in MeV) for 
          some fission channels   for the  $^{60}$Zn compound nucleus.} 

\begin{tabular}{|l|l|l|l|}
\hline
Reaction& Q-value & ${\hbar}^2/2\Theta_{ff}$ & Barrier($J$=45) \\
\hline
{\em Binary} (-- 0$\alpha$ ) & MeV &  & MeV \\
\hline
 $^{32}\textrm{S}$ + $^{28}\textrm{Si}$ & +3.34 & 0.0330  & 70.5 \\
 $^{30}\textrm{P}$ + $^{30}\textrm{P}$  & -3.76 & 0.0328  & 77.6 \\
\hline
{\em Binary} (-- 1$\alpha$) & MeV &  & MeV \\
\hline
$^{28}\textrm{Si}$+$^{28}\textrm{Si}$ + 1$\alpha$ & -3.60  & 0.0149 & 55.2  \\
$^{30}\textrm{P}$+$^{26}\textrm{Al}$ + 1$\alpha$ & -14.18  &   &   \\
\hline
{\em Ternary} (-- 2$\alpha$) & MeV &   & MeV\\
\hline
$^{24}\textrm{Mg}$+$^{28}\textrm{Si}$ + 2$\alpha$ & -13.58 &   &  68.5  \\
$^{26}\textrm{Al}$+$^{26}\textrm{Al}$ + 2$\alpha$ & -24.59 & 0.0136  &   79.4   \\
\hline
{\em Ternary} (-- 3$\alpha$) & MeV &   & MeV\\
\hline
$^{24}\textrm{Mg}$+$^{24}\textrm{Mg}$ + 3$\alpha$ & -23.57 & 0.0134  &   74.4  \\
$^{22}\textrm{Na}$+$^{26}\textrm{Al}$ + 3$\alpha$ & -34.04 &   &      \\
\hline
{\em Ternary} (-- 4$\alpha$) & MeV &   & MeV \\
\hline
$^{22}\textrm{Na}$+$^{22}\textrm{Na}$ + 4$\alpha$ & -43.5 & 0.0140  & 90.1   \\
$^{16}\textrm{O}$+$^{28}\textrm{Si}$ + 4$\alpha$  & -27.63 &   &     \\
\hline
\end{tabular}
\label{tab:1}
\end{table}

\begin{figure*}[htbp]
  \begin{center}
   \includegraphics[width=0.58\textwidth]{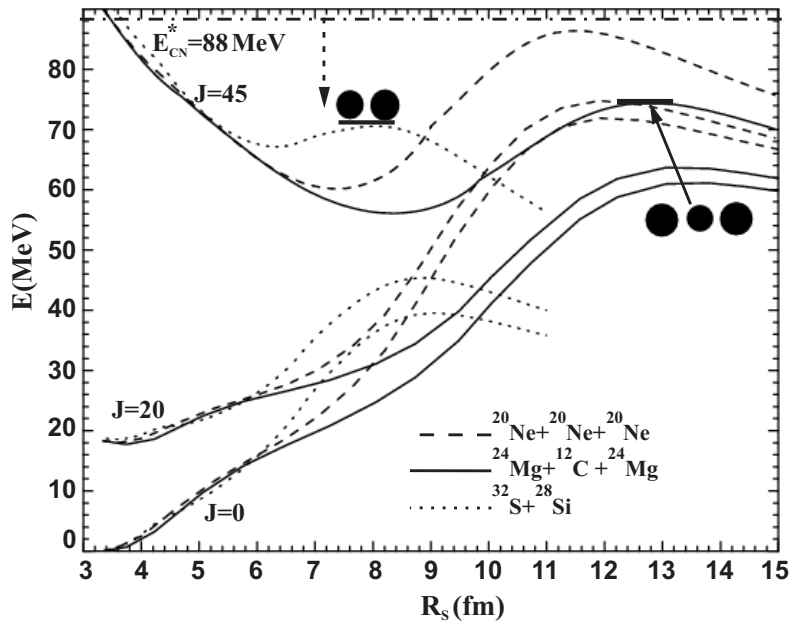}
    \caption{Potential energies obtained in the liquid drop model 
             for selected fragmentations in the 
            decay of $^{60}$Zn 
            as a function of the deformation (represented by the distance $R_s$
            between the two heavier fragments) for different angular
             momenta (in units of $\hbar$)
            for binary and ternary fission decay, respectively. 
            Channels with different  $\Delta$Z, (missing 0 and 
             3 $\alpha$-particles)  are shown.}
    \label{fig:potentialsArMg}
  \end{center}
\end{figure*}

In  Table~\ref{tab:1} we show  for different  binary and ternary
fission channels, both the ground state Q-values and the rotational energies at
the saddle point for $J = 45\hbar$, which will determine the yields.  For small
 values of $J$ the ternary fission probabilities are expected to be small, 
because they can not compete with the other decay channels. 
The Q-values for ternary mass splits attain large negative values.
The effective excitation energies 
of the primary fragments in a binary decay 
before statistical emission (e.g. for 3$\alpha$) must be rather higher, 
therefore the effective Q-values are much larger than those quoted in the table..

Concerning the experimental odd-even effect in the binary yields, we find that 
the Q-values for odd-odd charge fragments are typically by 5-10 MeV more negative.
In the same way the odd-even effect with the lower yields of the ternary
splits with odd charges  can   be understood, because less free energy is
available with the increasingly negative Q-values (up to Z$_{CN}$-6)
 for the odd-Z fragments.
These  smaller yields are in accordance with  the phase space considerations 
and the penetrabilities. In these cases also less excitation 
energy in the two heavier fragments is expected, and less subsequent decays via
particle evaporation. In fact the narrow peaks in the
($\phi_3-\phi_4$)-correlations dominate in the spectra if the sum of 
two odd charges (for $\Delta$Z=4, 6) is taken (see Fig.~\ref{fig:phiAr}).  

 The liquid-drop
energies  and the Q-values, as well as the rotational energy constitute the 
 main part of the barrier height for the fission process.
For the present reaction  these have been  calculated  
 by Royer~\cite{royer95}, where the liquid drop part of the potential energies 
is determined within a generalised liquid-drop model using quasi-molecular shapes. 
In the binary and ternary fission channels the saddle point corresponds to
 clusters kept together by the proximity forces. For the relative position of
 the binary and ternary fission barriers, clearly the ternary
fission barrier fgor   low values of $J$ is much higher, and no ternary fission can be observed.
 With the inclusion of the angular momentum,
 the rotational energy  as a function of  $J$ 
increases much slower at the strongly deformed saddle point of the ternary mass split
as shown in Fig.\ref{fig:potentialsArMg}. In this figure three cases are shown, a symmetric 
ternary mass split, a ternary channel with a  $^{12}$C-cluster 
in the centre and a binary channel.
The difference 
 relative to the binary mass split  becomes smaller because of the differences 
in the   moments of inertia, $\Theta$. From these calculations
 we also notice that for the ternary fission process the barrier is
the highest for three equal mass-fragments, whereas the barrier height
decreases if the mass (and charge) of the ``third'' fragment in the neck region
is smaller. Furthermore, the potential energy minima formed in the highly 
deformed rotating states are lower in the ternary valley as compared to 
the binary valley for large angular momenta.

  Following these considerations the
ternary fission process can be found to favourably compete with the binary mass
split at  the highest angular momenta, at values around 42-45 $\hbar$.
 An additional lowering
of the barrier by deformed-shell corrections is predicted 
for large deformations in many calculations based mainly on the
Nilsson-Strutinsky method~\cite{aberg90}. These particular shapes of the
hyper-deformation bring into play a deformed shell correction to the 
liquid-drop energy of approximately 5-10 MeV~\cite{Rag81}. The ternary fission
process from the hyper-deformed configuration can thus be further enhanced. 

We summarise our result  with the yield systematics shown in Fig.~\ref{fig:yieldsAr},
 where we can see a strong odd-even effect in both binary and ternary cluster decay
  (plotted as a function of the charge asymmetry).
  The experimental values are subject to some systematic errors,
 because the efficiency of
the BRS-coincidence set-up is only  optimally suited for symmetric mass splits 
and for  negative Q-values of at least $-$20 MeV. For the most asymmetric mass splits, 
and less negative  Q-values, the experimental yields (as deduced from
angle-angle, and energy-energy correlations) are sometimes reduced by 30$\%$,
 because of cuts in the geometrical angular range or in the energy range. 
This effect is partially incorporated into the systematic errors given for the
 individual yields. In accordance with the statistical model expectations, 
we observe a maximum in the yields for the case of one
missing  $\alpha$-particle.
In the binary mass split at least one of the fragments 
 has sufficient excitation energy for the  evaporation of  one 
 $\alpha$-particle, making the channel (Z$_{CN}$-2) the most
intense.
For  all very negative Q-values the yields are lower and the 
fragments are expected to be rather ``cold''. 

\section{Conclusions}
\label{concl}

We conclude that the  observation in the out-of-plane correlations 
of the very narrow coplanar fission 
fragment coincidences is a unique feature, observed for the first time, because of the 
unique detection efficiency of the BRS (see also ref.~\cite{voe08} for case of 
the fission decay of $^{56}$Ni. These data give 
clear evidence for the occurrence of  ternary cluster decay processes. 
From our analysis we conclude that, these must originate from the  strongly (hyper) 
deformed states in  nuclei with mass A=60 formed at high angular momentum as predicted 
by Zhang et al.~\cite{zhang94} as well as in the liquid drop model of Ref.~\cite{swiatecki}. 
 The  neck of the fissioning nucleus, as  calculated by Royer~\cite{royer95},
represents a region  of low nuclear density, which gives  an additional reason 
for the  favoured formation of  
$\alpha$-clusters, an  aspect which also has been discussed in the 
literature (see ref.~\cite{horiuchi}).

The systematics of the yields shown in Figs.~\ref{fig:yieldsAr} 
confirm the expectations of the statistical  model, i.e. that there is a strong 
odd-even effect in the yields(lower yields for the odd charges with more negative Q-values)
  for all cases of the binary 
and the ternary cluster decays.  A more detailed analysis,
 based on the mentioned statistical decay 
model is in preparation.

 The particles formed in 
 the neck are expected to travel with the centre-of-mass velocity in the
beam direction (towards 0$^\circ$). A corresponding measurement showing this
phenomenon, one  $\alpha$-particle from the neck in the decay of $^{28}$Si
into $^{12}$C+$\alpha$+$^{12}$C has been reported by Scheurer et al.~\cite{scheurer}. 
The present  work also shows that
the search for hyper-deformation in rapidly rotating nuclei, which has been done 
extensively using $\gamma$-spectroscopy, should be pursued 
with charged particle spectroscopy. For nuclei in the
medium-mass region (A = 40-100), where the fissibility parameter is small, 
the formation of a longer neck is predicted~\cite{swiatecki}. For  these cases ternary cluster 
decays are expected.
With three fragments  a complete reconstruction of the 
events can be undertaken with appropriate detector systemsm, which should  include a
zero degree hodoscope to detect the clusters formed in  the neck.
Such measurements offer the possibility of a precise spectroscopy of the
extremely deformed states.

Acknowledgments.
We thank H.G.~Bohlen, C.~Wheldon and Tz.~Kokalova for their numerous 
discussions on this project, H.G. Bohlen has contributed to many aspects of this work.
 V. Zherebchevsky thanks the DAAD for
a grant.


\begin{thebibliography}{99}

\bibitem{swiatecki}  S. Cohen, F. Plasil, and W. J. Swiatecki,
                     Ann. Phys. (N.Y.) \textbf{82} (1974) 557.

\bibitem{moeller76}  P. Moeller and R. Nix, 
                     Nucl. Phys. \textbf{A 272} (1976) 502.

\bibitem{zhang94}  J. Zhang, A. C. Merchant, and W.D.M. Rae,
                 Phys. Rev. \textbf{C 49} (1994) 562; 
                 see also W.D. M. Rae in Proc.,{\it 5$^{th}$ Intern.
                 Conf. on Clustering Aspects in Nuclear and Subnuclear Systems}
               (1988), Kyoto, Prog. Theor. Phys. (Jap.) ed. K. Ikeda, 1989, p.80.

\bibitem{lea75}  G.~Leander and S.E.~Larsson, 
                 Nucl. Phys. {\bf A239} (1975) 93.

\bibitem{aberg90} S. Aberg, H. Flocard, and W. Nazarewicz, 
                  Ann. Rev. Nucl. Science. {\bf 40} (1990) 439.

\bibitem{aberg94} S. Aberg and L. O. Joensson, 
                  Z. Phys. {\bf A 349} (1994) 205.

\bibitem{Rag81} I. Ragnarsson, S. Aberg, and R. K. Sheline,
                Physica Scripta {\bf 24}, 215 (1981);
                I. Ragnarsson, S.G. Nilsson, and R.K. Sheline, 
                Phys. Rep. {\bf 45} (1978) 1. 

 \bibitem{sanders94}  S.J. Sanders, et al., 
                      Phys. Rev. \textbf{C 49} (1994) 1016, and refs. therein.

\bibitem{sanders87}
                S.J. Sanders, D.G. Kovar, B.B. Back, C. Beck, B.K. Dichter,
                D. Henderson, R.V.F. Janssens, J.G. Keller, S. Kaufman, 
                T.-F. Wang, B. Wilkins, and F.Videbaek, 
                Phys. Rev. Lett. {\bf 59}, (1987)  2856.

\bibitem{sanders89} 
                    S.J. Sanders, D.G. Kovar, B.B. Back, C. Beck, D.J.
                    Henderson, R.V.F.
                    Janssens, T.F. Wang, and B.D. Wilkins, 
                    Phys. Rev. C {\bf 40}, (1989)  2091.

\bibitem{sanders99}   S.J. Sanders, A. Szanto de Toledo, and C. Beck,
                      Phys. Rep. \textbf{311} (1999) 487, and references therein.



\bibitem{kokalova05}  Tz.~Kokalova, W. von Oertzen, S. Torilov et al.,
                      Eur. Phys. J. {\bf A 22 } (2006) 19.

\bibitem{Andreev06}  A. V. Andreev, G. G. Adamian, N. V. Antonenko  et al.,
                     Eur. Phys. J.  \textbf{A 30} (2006) 579.

\bibitem{roy95} G. Royer and F. Haddad,
                J. Phys. \textbf{G 21} (1995) 339.


\bibitem{royer95} G. Royer,
                   J. Phys. \textbf{G 21} (1995) 249; 
                  see also G. Royer, F. Haddad, and J.Mignen,
                  J. Phys. \textbf{G 18} (1992) 2015. 



\bibitem{murphy96} A. St. J.~Murphy, et al.,
                   Phys. Rev. C \bf 5\rm (1996) 1963.

\bibitem{thummerer98} S. Thummerer et al., 
                      Nuovo Cimento \textbf{111 A} (1998) 1077, 
                      and Dr.Thesis, Freie Universitaet, Berlin, 1999.

\bibitem{zhereb06}  V. I. Zherebchevsky, W. von Oertzen, D. Kamanin, et al.
                   Phys. Lett. \textbf{B 646} (2007) 12.
 
\bibitem{zhereb07}  V. I.  Zherebchevsky, W . von Oertzen, D. V.  Kamanin, 
                     JETP Letters  Vol. \textbf{85}, N.3, (2007) 136 and 
                    V.I. Zherebchevsky, Dr.Thesis, St. Petersburg University (2007)

\bibitem{Efimov07} G. Efimov et al., 
                    Dr. Thesis, Dubna University (2008) 

\bibitem{voe08}    W.  von Oertzen, B. Gebauer, G. Efimov et al., 
                   Eur. Phys. J. A \textbf{36} (2008) 279    

\bibitem{beck96}   C.~Beck and A. Szanto de Toledo,
                   Phys. Rev. C \textbf{53} (1996) 1989.

\bibitem{ref:gebauer98} B.~Gebauer et al., in \it Ancillary Detectors 
                        and Devices for Euroball\rm, 1998, ed. H. Grawe, p.43; 
                        see also \it Achievements with the Euroball
                        spectrometer\rm, 1997-2003,
                        ed. W. Korten and S. Lunardi, p. 135. 

\bibitem{beck04} C. Beck et al., Nucl. Phys. \textbf{A734} (2004) 453.

\bibitem{ref:gebauer93} B.~Gebauer et al., in Proc. Int. Conf. on the 
                        Future of Nucl. Spectroscopy, Crete, Greece, 
                        1993, eds.W. Gelletly et al., p. 168.



\bibitem{matsuse97} T. Matsuse, C. Beck, R. Nouicer, and D. Mahboub, 
                    Phys. Rev. C {\bf 55} (1997) 1380.


\bibitem{morgenstern} H. Morgenstern et al.,
                      Z. Phys. \textbf{A 313} (1983) 39.



\bibitem{wuosmaa89} A.H. Wuosmaa et al.,
                    Phys. Rev. C {\bf 40} (1989) 173.

\bibitem{zhukov81} H. J. Wiebecke and M. Zhukov, Nucl. Phys. 
                   Nucl. Phys. \textbf{A 351} (1981) 321.

\bibitem{horiuchi}  H. Horiuchi, Nucl. Phys. \textbf{A731} (2004) 329.

\bibitem{scheurer}  J.N. Scheurer et al., 
                    Nucl. Phys. \textbf{A 319} (1979) 274.


\end{thebibliography}
\end{document}